\documentclass[10pt,journal,compsoc]{IEEEtran}

\IEEEoverridecommandlockouts
\ifCLASSOPTIONcompsoc
  \usepackage[nocompress]{cite}
\else
  \usepackage{cite}
\fi
\usepackage{orcidlink}
\usepackage{enumitem}
\usepackage{caption}
\usepackage{amsthm,amsmath,amssymb,amsfonts,mathrsfs}
\usepackage{bbm}
\usepackage[noend]{algorithmic}
\usepackage{algorithm}
\usepackage{subfigure}
\usepackage{graphicx}
\usepackage{textcomp}
\usepackage{xcolor}
\usepackage{dsfont}
\usepackage{float}
\usepackage{tabularx}
\usepackage{makecell}
\usepackage{array}
\usepackage{colortbl}
\usepackage{siunitx}
\usepackage{tikz}
\newcolumntype{?}{!{\vrule width 1pt}}

\newcommand{\eq}[1]{Eq.~\eqref{#1}}
\newcommand{\beq}{\begin{equation}}
\newcommand{\eeq}{\end{equation}}
\newcommand{\bit}{\begin{itemize}}
\newcommand{\eit}{\end{itemize}}

\usepackage{multirow}
\usepackage{url}
\usepackage{bm}
\renewcommand{\baselinestretch}{0.988}
\usepackage{booktabs}
\def\BibTeX{{\rm B\kern-.05em{\sc i\kern-.025em b}\kern-.08em
    T\kern-.1667em\lower.7ex\hbox{E}\kern-.125emX}}
\usepackage{hyperref} 
\hypersetup{
    linkbordercolor={0 0 1},
    citebordercolor={0 0 1},
    filebordercolor={0 0 1}
}
\usepackage{mdframed}
\mdfdefinestyle{ProblemFrame}{%
    linecolor=black,
    outerlinewidth=2pt,
    innertopmargin=2pt,
    innerbottommargin=4pt,
    innerrightmargin=4pt,
    innerleftmargin=4pt,
        leftmargin = 4pt,
        rightmargin = 4pt
    }

\newcommand\diagfil[4]{%
  \multicolumn{1}{p{#1}}{\hskip-\tabcolsep
  $\vcenter{\begin{tikzpicture}[baseline=0,anchor=south west,inner sep=0pt,outer sep=0pt]
  \path[use as bounding box] (0,0) rectangle (#1+2\tabcolsep,\baselineskip);
  \node[minimum width={#1+2\tabcolsep},minimum height=\baselineskip+\extrarowheight+0+0,fill=#2] (box)at(0,0) {};
  \fill [#3] (box.south west)--(box.north east)|- cycle;
  \node[anchor=center] at (box.center) {#4};
  \end{tikzpicture}}$\hskip-\tabcolsep}}
  
\newcommand{\argmin}{\mathop{\mathrm{argmin}}}    

\makeatletter

\def\ps@IEEEtitlepagestyle{%
\def\@oddhead{\parbox[t][\height][t]{\textwidth}{\centering
Published in IEEE Transactions on Mobile Computing (Open Access): \url{https://doi.org/10.1109/TMC.2022.3185681}.\\
}\hfil\hbox{}}%
\def\@evenhead{\parbox[t][\height][t]{\textwidth}{\centering
Published in IEEE Transactions on Mobile Computing (Open Access): \url{https://doi.org/10.1109/TMC.2022.3185681}.\\
}\hfil\hbox{}}}
\def\@oddhead{\parbox[t][\height][t]{\textwidth}{\centering
Published in IEEE Transactions on Mobile Computing (Open Access): \url{https://doi.org/10.1109/TMC.2022.3185681}.\\
}\hfil\hbox{}}%
\def\@evenhead{\parbox[t][\height][t]{\textwidth}{\centering
Published in IEEE Transactions on Mobile Computing (Open Access): \url{https://doi.org/10.1109/TMC.2022.3185681}.\\
}\hfil\hbox{}}

\makeatother

\begin{document}

\title{SHARP: Environment and Person Independent\\Activity Recognition with Commodity\\IEEE 802.11 Access Points}

\author{Francesca Meneghello\orcidlink{0000-0002-9905-0360}, Domenico Garlisi\orcidlink{0000-0001-6256-2752}, Nicol\`o Dal Fabbro\orcidlink{0000-0002-5325-2792}, Ilenia Tinnirello\orcidlink{0000-0002-1305-0248}, Michele Rossi\orcidlink{0000-0003-1121-324X}
\IEEEcompsocitemizethanks{
\IEEEcompsocthanksitem F. Meneghello, N. Dal Fabbro and M. Rossi are with the Department of Information Engineering (DEI), University of Padova, via Gradenigo 6/b, 35131, Padova, Italy. M. Rossi is also with the Department of Mathematics ``Tullio Levi-Civita'', via Trieste 63, 35131, Padova, Italy. F. Meneghello is the corresponding author. E-mail: \{name.surname\}@dei.unipd.it
\IEEEcompsocthanksitem D. Garlisi and I. Tinnirello are with the Department of Engineering, University of Palermo, Viale delle Scienze, edificio 7, 90128, Palermo, Italy and also with CNIT, 43124 Parma. E-mail: \{name.surname\}@unipa.it
}}

\IEEEtitleabstractindextext{
\begin{abstract}
In this article we present SHARP, an original approach for obtaining human activity recognition (HAR) through the use of commercial IEEE 802.11 (\mbox{Wi-Fi}) devices. SHARP grants the possibility to discern the activities of different persons, across different time-spans and environments. 
To achieve this, we devise a new technique to clean and process the channel frequency response (CFR) phase of the \mbox{Wi-Fi} channel, obtaining an estimate of the Doppler shift at a radio monitor device.
The Doppler shift reveals the presence of moving scatterers in the environment, while not being affected by (environment-specific) static objects. 
SHARP is trained on data collected as a person performs seven different activities in a single environment. It is then tested on different setups, to assess its performance as the person, the day and/or the environment change with respect to those considered at training time. In the worst-case scenario, it reaches an average accuracy higher than $95\%$, validating the effectiveness of the extracted Doppler information, used in conjunction with a learning algorithm based on a neural network, in recognizing human activities in a subject and environment independent way. The collected CFR dataset and the code are publicly available for replicability and benchmarking purposes~\cite{dataset-code}.
\end{abstract}
}

\maketitle
\IEEEdisplaynontitleabstractindextext

\IEEEraisesectionheading{\section{Introduction}}
Human activity recognition (HAR) systems are key elements of emerging applications devised for implementation in smart buildings. Among their other functionalities, they can be profitably exploited to optimize the buildings’ energy consumption, implement alert systems, provide platforms for smart entertainment, etc. They can also be used to complement IT solutions for elderly care, to promptly detect critical situations or, for assisted-living purposes, to support physically-impaired people in their daily activities~\cite{Dong2019}. 
HAR can be performed in indoor environments through ad-hoc devices -- such as radars and cameras -- or by leveraging already deployed systems without the burden of installing additional instruments. This second possibility is being investigated by several researchers and has recently attracted the interest of the industrial community. In this context, in September 2020, the IEEE established a working group for the definition of a new version of the IEEE 802.11 (Wi-Fi) standard, named IEEE 802.11bf. The new specifications will include modifications at the physical and medium access control layers of the protocol stack to enable the joint provisioning of the communication and the human/environmental sensing services~\cite{Restuccia2021}. Devices implementing the new standard are expected to be commercialized by 2024. However, the standard is focused on innovative and cooperative control mechanisms for triggering and exchanging sensing measurements. The definition of algorithms for processing these measurements and detect features of a target, under unknown environmental conditions, is still an open issue.

In the last decade, since the pioneering work of Adib and Katabi~\cite{Katabi}, wireless signals of opportunity have been extensively researched as a means to perform \mbox{device-free} localization and activity recognition tasks, including pattern and gesture recognition~\cite{Zheng2019}, and even for estimating biological signals at distance, such as the respiration rate or the heartbeat~\cite{Wang2018}. In this respect, \mbox{Wi-Fi} signals are particularly appealing, due to the high availability of \mbox{Wi-Fi}-enabled devices in most residential and working spaces~\cite{He2020}. 
While active sensing techniques allow localizing and following users as they carry along a \mbox{\mbox{Wi-Fi}-connected} device, passive sensing approaches obtain such information by monitoring the changes in the channel, where users act as scatterers that modify the channel frequency response (CFR), a frequency estimate of the radio channel describing how the signal changes as it propagates from the transmitter to the receiver~\cite{Yongsen2019}. 
The CFR -- also referred to as the channel state information (CSI) -- is continuously estimated by \mbox{Wi-Fi} routers for communication purposes. Recently, specific tools have allowed gathering such information from commercial off-the-shelf (COTS) devices, enabling its use for environmental sensing applications.

In this work, we present and validate SHARP (\textit{sensing human activities through \mbox{Wi-Fi} radio propagation}), a device-free HAR system for the automatic detection and classification of human activities in indoor spaces. 
SHARP is a novel approach to HAR that leverages standard-compliant Wi-Fi devices and achieves unobtrusive human sensing through the analysis of the CFR gathered by COTS IEEE~802.11 routers. We emphasise that, although similar sensing technology has already been proposed as part of prior work, e.g., ~\cite{Katabi, Zheng2019, Wang2018, He2020, Yongsen2019}, existing algorithms are unable, unlike SHARP, to generalize across unknown environments and people, and situations for which they were not trained. That is, they {\it fall short in generalizing across persons and environments}.
The proposed system sharply differs from these previous solutions, and improves upon them, in that it reliably works across different indoor environments and people, without having to be re-trained each time the physical wireless propagation setup, or the monitored person, change. To make this possible, SHARP features an integrated procedure combining some preliminary processing steps and a learning-based activity classification algorithm. At first, {\it an original phase-cleaning method} permits the extraction of {\it micro-Doppler} traces from the CFR data. Without this procedure, the phase offsets that affect any Wi-Fi transmission would be so strong as to hide finer phase variations and, in turn, would prevent the extraction of the micro-Doppler information. The proposed phase sanitization approach is a major step forward in the HAR via Wi-Fi channels: as we show in Section~\ref{sec:phase_sanitization_results}, prior sanitization algorithms fail to provide a sufficiently accurate phase signal and are thus ineffective to our purpose. Once available, the Doppler shift reveals the velocities of the scattering points during the transmission events and is not affected by static objects (furniture, walls, etc.), allowing the gauging of dynamic HAR features. As quantified in Section~\ref{sec:results}, we found it to be a robust representation for environment- and person-independent HAR. 
To do that, the micro-Doppler traces are used as input for a neural network architecture that is trained to recognize the activities of interest. The effective combination of the preliminary processing steps and the learning algorithm ensures high performance also in changing environmental conditions. This, of course, is a much-desirable feature in off-the-shelf implementation for smart-home applications.

Our work focuses on the recognition of dynamic activities performed by a human in an indoor environment. The identification of static activities -- such as relax poses -- is out of the scope of the present work. 
For the recognition of static activities, a paradigm change in both the monitoring technology and the data processing would be needed. Using \mbox{Wi-Fi} devices operating in the sub-$6$~GHz bands, static users can be hardly distinguished from the empty room situation as the features that can be acquired or computed, e.g., amplitude, phase, Doppler, can not reliably reveal the presence of a static human. In case the amplitude information is considered, the static person only introduces an offset that is constant over time and which would be hardly distinguishable from small changes in the positions of objects inside the room (as discussed at the beginning of Section~\ref{subsec:doppler_computation}). Moreover, the presence of a static person has no impact on the Doppler traces as no micro-Doppler effect is introduced. The recognition of static poses through radio-wave technologies would be possible by leveraging devices operating at higher frequencies, e.g., millimeter-wave radars~\cite{Cui2022real, Sengupta2022mmpose}. However, the detection of static poses is less relevant than dynamic activities in the development of smart-home applications, e.g., involving the contactless monitoring of people. In this context, the main objective is the detection of dynamic movements that may reveal normal or critical situations. SHARP provides proper support for these use cases through standard-compliant IEEE 802.11ac devices that nowadays are almost ubiquitous in residential and working spaces.

The present work sharply departs from the literature on HAR with \mbox{Wi-Fi} devices, as follows.

\noindent$\bullet$ We propose an original method for CSI phase correction (also referred to as \textit{sanitization}). The algorithm is independently used at each receiving antenna without the need for a reference stream. Hence, the space diversity provided by the antennas can be fully exploited for sensing purposes.

\noindent$\bullet$ We present and validate SHARP, an environment- and person-independent learning-based framework for HAR, which leverages the Doppler effect caused by human motion. The information collected at the monitoring antennas is combined to identify the activity, regardless of the user's position in the monitored area. 

\noindent$\bullet$ We assess SHARP robustness to changing environmental conditions through CSI data collected with commercial off-the-shelf IEEE 802.11ac devices operating on an $80$~MHz frequency band. It reaches an average accuracy of $96\%$ in the identification of four activities -- namely walking, running, jumping and sitting -- in the most challenging scenario, i.e., when the person, the day and the indoor environment change with respect to those provided in the training set. SHARP is validated against prior work -- DeepSense~\cite{Zou2018}, EI~\cite{TOWARDS} and MatNet-eCSI~\cite{Shi2020} -- achieving superior generalization capabilities across subjects and environments.
SHARP can be extended to recognize more activities in an environment-independent manner. The numerical results show that it allows reaching an accuracy higher than $95\%$ in the recognition of seven activities, being standing, doing arm gymnastics and sitting down/standing up in addition to the former ones.
Note that we validate SHARP on devices implementing the most commonly used \mbox{Wi-Fi} standard at the time of writing, namely the IEEE 802.11ac. However, our approach can also be integrated -- without modification -- with older and upcoming IEEE standards adopting the orthogonal frequency division multiplexing (OFDM) modulation scheme, such as IEEE 802.11n/ax/be/bf. We released the dataset of the collected CFR traces and the code implementing the devised algorithm to allow replicability and provide a performance benchmark~\cite{dataset-code}.

The rest of the paper is organized as follows. The related work is reviewed in Section~\ref{sec:related}. In Section~\ref{sec:processing} and Section~\ref{sec:learning}, we respectively detail the processing on the CSI data and the learning architecture for HAR, which represent the SHARP building blocks. The experimental setup is presented in Section~\ref{sec:setup}, while the performance of the proposed approach is evaluated in Section~\ref{sec:results}. Conclusions are drawn in Section~\ref{sec:conclusions}. For completeness, we include Appendix~A where an overview of the OFDM \mbox{Wi-Fi} channel model is provided.\vspace{-0.2cm}

\section{Related work}\label{sec:related}
In this section, to put our contribution into context, we review the existing literature on \mbox{\mbox{Wi-Fi}-based} HAR. We specifically focus on works tackling the robustness and generalization of HAR algorithms to different days, environments, and people. Hence, we review the contributions regarding the sanitization of the CFR phase information, the extraction and use of micro-Doppler features.
\subsection{CSI based human activity recognition}\label{subsec:act_rec}
HAR through CSI data from COTS \mbox{Wi-Fi} devices was first studied in, e.g., E-eyes~\cite{EYES}, CARM~\cite{Wang2015}.
More recently, several articles showed the effectiveness of machine learning techniques in building algorithms that distinguish human activities based on CSI features~\cite{Zou2018, Ding2019, Damodaran2020, Tegou2020}. 
However, these works do not focus on the robustness to environmental changes and on the generalization capability to previously unseen environments and subjects, which are key enablers for the successful development of \mbox{Wi-Fi}-based sensing systems~\cite{Yongsen2019}. 
Only a few works try to address these weaknesses. In~\cite{TOWARDS}, the authors use a neural network-based approach to extract environment-independent features from the CSI amplitude to recognize human movements. The performance of their algorithm is promising, but remains below $80\%$ in the best scenario. In~\cite{TRUESCALING}, transfer learning is shown to be effective to adapt the \mbox{Wi-Fi}-based HAR algorithm to different persons and days for the same environment. The algorithm presented in~\cite{Xiao2019} leverages generative adversarial networks to generalize on new persons, while in~\cite{Shi2020} the matching network one-shot learning approach~\cite{MATNET} is proposed to bridge the gap between previously seen environments and new ones. 
A recent work~\cite{LOCATION} addresses the problem of location and subject independent HAR through a learning architecture consisting of three deep neural networks. The algorithm is trained on the CSI amplitude collected by monitor routers placed in different positions inside a room, and is tested in the same room by changing the location of a single router. The approach presents a significant performance degradation when evaluated on other datasets (going from $99\%$ to $80\%$ of accuracy).

To the best of our knowledge, no work in the literature proposes a system that generalizes well on unknown environments, days and subjects without any re-training step and using COTS devices. 
In the present work, we propose an effective solution to this problem, and we consider as benchmarks for comparison, three recent approaches from the literature, i.e., DeepSense~\cite{Zou2018}, EI~\cite{TOWARDS}, MatNet-eCSI~\cite{Shi2020}.\vspace{-0.3cm}

\subsection{Exploitation of the CFR phase}\label{subsec:phase_sanit}
Some works, e.g,~\cite{WangX2020, Zeng2018, Zhang2019}, consider a reference antenna and use the phase differences at the remaining ones to perform sensing. By construction, such phase difference is not affected by systematic offsets (same value across the antennas), but it is unable to cope with offsets that are \mbox{antenna-specific}.
Exploiting a similar idea, in~\cite{Keerativoranan2018}, the authors correct the rotation errors using a reference signal, obtained by connecting with a cable one of the available monitoring antennas with the transmitter. 
However, the need for a reference reduces the spatial diversity that can be exploited  for sensing purposes at the monitor station.
The possibility to remove the unwanted phase offsets without exploiting any reference is less addressed in the literature. In~\cite{Borhani2019}, the authors propose to mitigate the errors by considering the average signal over a number of time instants, while in~\cite{Chen2019} a two-step approach is presented to estimate and remove some of the \mbox{antenna-independent} offsets. A different strategy is adopted in \cite{Shi2020}, where the authors estimate the systematic phase shift at each CFR acquisition. With their approach they remove the static component, that is used as a reference for the algorithm, and retain the dynamic one, using it as the input for the subsequent HAR framework.

In this work, we correct the phase offsets by devising an optimization approach that finds the optimal CFR parameters from the raw channel data. For each CFR sample, we extract the contributions of the different radio paths and exploit the component related to the strongest path as a reference to correct the signal phase. This allows the removal of the non-systematic offsets that affect the signal. As discussed above, this would result infeasible if considering as a reference a signal extracted form a different physical measurement (either in the time or in the spatial dimension).\vspace{-0.3cm}

\subsection{Doppler based applications}\label{subsec:doppler_appl}
The micro-Doppler effect extracted from ad-hoc RADAR devices has been extensively used in the literature to perform HAR. Several studies emphasise its effectiveness in highlighting relevant features for the correct identification of the dynamics in the scene~\cite{Pegoraro2020}. However, only a few works in the literature exploit the Doppler shift computed from commercial \mbox{Wi-Fi} devices, without recurring to ad-hoc infrastructures. In~\cite{Li2017, Qian2017, Qian2017_2}, the Doppler information is used to track the movement of a user inside an environment. Two different monitoring devices, placed in strategic positions, are used to obtain effective Doppler shift estimates.
In~\cite{Wenda2021passive} the authors leverage the Doppler effect for breathing detection and people counting. The sensing system consists of one Wi-Fi access point and two universal software radio peripherals (USRPs) monitoring the target channel and a reference one that needs to remain stable.
A sensing system consisting of a single monitoring station is presented in~\cite{Li2018}, to recognize three different bodyweight exercises. In~\cite{Li2018}, the transmitter and the monitor stations must be placed in specific positions with respect to the user, so that the activity is performed perpendicularly to the radio link. Moreover, the detection algorithm is based on metrics that are manually extracted from the raw CSI data and the Doppler trace. This makes the proposed approach highly \mbox{activity-dependent}, as the features are specifically designed to separate the three considered physical exercises. HAR through Doppler information has been recently addressed in~\cite{Muaaz2020}.
The latter algorithm uses phase sanitization based on a reference antenna, followed by a support vector machine (SVM) classifier.
This strategy allows achieving a good recognition accuracy ($96\%$) on test data collected in the same environment where the algorithm is trained, proving the effectiveness of the Doppler information for HAR purposes. 
Also, the system performance is improved considering additional data from a wearable inertial measurement unit (IMU), increasing the accuracy to nearly $100\%$. However, no performance assessment is carried out for scenarios that are not seen at training time. 
SHARP outperforms \cite{Muaaz2020} by exploiting an advanced phase sanitization approach and a more powerful learning architecture. As a result, it reaches accuracies close to $100\%$ by only exploiting CSI data (no IMU). Moreover, SHARP generalizes to unknown environments, by correctly recognizing the activity in at least $95.99\%$ of the cases.

Our intuition, behind the design of SHARP, is that the Doppler shift naturally lends itself to the separation of the radio reflections coming from static objects or structures (e.g., room walls) from those generated by moving targets (e.g., humans). If available, it can be used as a reference signal domain to perform environment independent recognition tasks, assuming that the environment is mostly static. Our work descends from this line of reasoning and our novelty resides in the fact that we were, for the first time, able to extract the Doppler signal from $802.11$ac COTS devices. After that, such Doppler signal must be used in combination with dedicated processing and learning architectures, which are the second main contribution of our present work. In the remainder of this work, the extraction of the Doppler is presented in detail, along with its use with inception-based neural networks.\vspace{-0.4cm}

\section{CSI data processing}\label{sec:processing}
\mbox{Wi-Fi} systems adopt orthogonal frequency division multi-plexing (OFDM), by transmitting the user information over $K$ partially overlapping and orthogonal \mbox{sub-channels}, with $K$ even. At the OFDM receiver, the channel parameters (amplitudes and phases) are continuously estimated for all the \mbox{sub-channels}. This information is computed for each received packet based on known preamble symbols, and is collected by the CSI, a large (environment-dependent) complex matrix describing the CFR for each \mbox{sub-channel} along every receiving antenna.\footnote{In the rest of the paper, the terms CSI and CFR will be used interchangeably.}

For each pair of transmit and receive antennas, the CFR is a set of complex numbers $A_k e^{j\phi_k}$ specifying the attenuation $A_k$ and the phase shift $\phi_k$ for each \mbox{sub-channel} \mbox{$k \in \{-K/2, \dots, K/2 - 1\}$}. Considering a \mbox{multi-path} propagation channel, and using index $p \in \{0, \dots, P-1\}$ to indicate the $P$ copies of the transmitted signal that are collected at the receiver, the CFR estimated on packet $n$ is written as
\beq \label{eq:completeh_clean}
\begin{split}
	 H_{k}(n) &= A_k(n) e^{j\phi_k(n)} \\
	 &=\sum_{p=0}^{P-1} A_{p}(n) e^{-j2\pi (f_c + k/T)\tau_{p}(n)},
\end{split}
\eeq
where $f_c$ is the main carrier frequency, $T=1/\Delta f$ is the OFDM symbol time, with $\Delta f$ being the sub-channel spacing (see Appendix~A), while $A_{p}(n)$ and $\tau_{p}(n)$ respectively represent the attenuation and the delay associated with path $p$. The complete OFDM model for the \mbox{Wi-Fi} channel, along with a derivation of \eq{eq:completeh_clean} are given in Appendix~A.

In this work, the CFR matrix is estimated by a monitor \mbox{Wi-Fi} device running the Nexmon CSI tool~\cite{Schulz2017}, as detailed in Section \ref{sec:setup}. We remark that the collected CFR slightly deviates from the theoretical model in \eq{eq:completeh_clean} due to hardware artifacts, which introduce an undesired phase offset $\phi_{{\rm offs}, k}$, i.e., 
\beq \label{eq:completeh}
	\bar{H}_{k}(n) = H_{k}(n) e^{\phi_{{\rm offs}, k}}.
\eeq

Next, we present the steps that we implemented to clean the CFR matrix that is extracted by the Nexmon CSI tool. Note that a new CFR matrix is retrieved for each received packet. Thus, the interval between subsequent acquisitions of the CFR is variable but, for the sake of exposition, in the following analysis we assume that a new sample is available every $T_c$ seconds, with $T_c$ fixed. 
We remark that, in a real \mbox{Wi-Fi} network, the traffic exchanged between already deployed access points and user devices (e.g., streaming services), can be captured and exploited for HAR recognition purposes. When this is not possible, two \mbox{Wi-Fi} access points are required to recreate the setup, using a simple network controller to transmit packets at regular intervals and a second monitor device to infer the CFR from these.\vspace{-0.3cm}

\subsection{Phase sanitization}\label{subsec:sanitization}
The undesired phase offset $\phi_{{\rm offs}, k}$ in \eq{eq:completeh} contains different contributions (see Appendix~A for their detailed explanation)~\cite{Zhu2018, Keerativoranan2018}. Some of them, namely, the channel frequency offset (CFO), the phase-locked loop (PPO) and the phase ambiguity (PA), although changing in time have the same value across the \mbox{sub-channels} in each receiving antenna. The sampling frequency offset (SFO) and the packet detection delay (PDD) are instead \mbox{sub-channel} dependent. As shown in Appendix~A, 
the offset $\phi_{{\rm offs}, k}$ experienced at one receiving antenna in \mbox{sub-channel} $k$ can be expressed as
\beq \label{eq:completephase_m}
	\phi_{{\rm offs}, k} = - 2\pi k(\tau_{{\rm SFO}} + \tau_{{\rm PDD}})/T  + \phi_{\rm CFO} + \phi_{{\rm PPO}} + \phi_{{\rm PA}}.
\eeq
The main idea behind our approach to phase sanitization is that the contribution of each path in \eq{eq:completeh} is affected by the same phase shift $\phi_{{\rm offs}, k}$. Hence, if we were able to separate the different paths, we could use (any) one of them as a reference to remove the phase offset from the CFR. In our method, the strongest path will be used to this end, as this is the path whose parameters are more reliably estimated at the receiver. The details are given shortly below. (Note that in the following analysis the time index $n$ is omitted in the interest of readability.)

Let $\mathbf{h}$ be the $K$-dimensional vector collecting the CFR information for the $K$ sub-channels, 
\beq
	\mathbf{h} = [\bar{H}_{-K/2}, \dots, \bar{H}_{0}, \dots \bar{H}_{K/2-1}]^T.
\eeq
To separate the $P$ \mbox{multi-path} contributions, we define a grid of $P'$ possible paths with $P' > P$ and solve a minimization problem to select the $P$ components out of the $P'$ ones that contribute to the CFR. We consider the following decomposition
\beq \label{eq:Hsplit}
	\mathbf{h} = \mathbf{T} \mathbf{r},
\eeq
where $\mathbf{T}=[\mathbf{T}_{-K/2} \dots \mathbf{T}_{K/2-1}]^T$ is an ($K\times P'$)-dimensional matrix collecting the contributions to the CFR that depend on the \mbox{sub-channel} index $k$, while $\mathbf{r}$ is a $P'$-dimensional column vector representing the \mbox{sub-channel} independent terms. 
The \mbox{$P'$-dimensional} row vectors $\mathbf{T}_k$ are defined through a dictionary of candidate total delays $\tau_{p, {\rm tot}} = \tau_p + \tau_{\rm SFO} + \tau_{\rm PDD}$, with $p \in \{0, \dots, P'-1\}$ as
\beq \label{eq:T_matrix}
	\mathbf{T}_k =
	\begin{bmatrix}
	e^{-j2\pi k \tau_{0, {\rm tot}}/T} & \dots & e^{-j2\pi k \tau_{P'-1, {\rm tot}} /T} 
	\end{bmatrix},
\eeq
The \mbox{$P'$-dimensional} column vector $\mathbf{r}$ can be modeled as  
\beq \label{eq:r_model}
	\mathbf{r}=e^{j(\phi_{\rm CFO} +\phi_{{\rm PPO}} + \phi_{{\rm PA}})}
	\begin{bmatrix}
	A_0 e^{-j2\pi f_c\tau_0}   \\
	\vdots &   \\
	A_{P'-1} e^{-j2\pi f_c\tau_{P'-1}} \\
	\end{bmatrix},
\eeq
and is obtained by solving the minimization problem

\begin{mdframed}[style=ProblemFrame,nobreak=true]
\begin{align}
    \label{eq:problem_P1}
    \textbf{P1}:~~~~
	\mathbf{r} = \argmin_{\tilde{\mathbf{r}}} ~ \left| \left| \mathbf{h} - \mathbf{T}  \tilde{\mathbf{r}}\right| \right|_2^2 + \lambda \left| \left| \tilde{\mathbf{r}} \right| \right| _1,
\end{align}
\end{mdframed}
where $\lambda >0$ is the weighing parameter for the $\ell_1$ regularization term in the Lasso regression. As $\mathbf{r}$ reflects the sparse channel impulse response,  \textbf{P1} is a compressive sensing reconstruction problem and can be solved through quadratic optimization using, e.g.,~\cite{osqp}.
Note that $\mathbf{h}$, $\mathbf{T}$ and, in turn, $\mathbf{r}$, are complex-valued vectors/matrices while the majority of quadratic optimization tools, e.g.,~\cite{osqp}, work with real numbers. In turn, to solve \textbf{P1}, we devised a strategy to convert our problem into a real-valued minimization task as follows. The \mbox{$K$-dimensional} complex-valued $\mathbf{h}$ vector is first converted into a \mbox{$2K$-dimensional} real-valued vector -- $\mathbf{h}_{\rm ext}$ -- by concatenating the real and the additive inverse of the imaginary parts, i.e.,
\beq 
	 \mathbf{h}_{\rm ext} =
	 \begin{bmatrix}
	 \operatorname{Re}\left(\mathbf{h}\right) & -\operatorname{Im}\left(\mathbf{h}\right) 
	 \end{bmatrix}.
\eeq
The \mbox{$K \times P'$-dimensional} complex-valued $T$ matrix is converted into a \mbox{$2K \times 2P'$-dimensional} real-valued matrix -- $\mathbf{T}_{\rm ext}$ -- by combining the imaginary and real parts as
\beq 
	 \mathbf{T}_{\rm ext} =
	 \begin{bmatrix}
	 \operatorname{Re}\left(\mathbf{T}\right) & -\operatorname{Im}\left(\mathbf{T}\right) \\
	 \operatorname{Im}\left(\mathbf{T}\right) & \operatorname{Re}\left(\mathbf{T}\right)
	 \end{bmatrix}.
\eeq
Using these conversions, $\mathbf{P1}$ in \eq{eq:problem_P1} is solved through \cite{osqp} obtaining a \mbox{$2P'$-dimensional} real-valued $\mathbf{r}_{\rm ext}$ vector. To obtain the \mbox{$P'$-dimensional} complex-valued $\mathbf{r}$ vector, the real and imaginary parts of $\mathbf{r}_{\rm ext}$ are recombined by applying the reverse of the processing explained above for $\mathbf{h}$, i.e., 
\beq
	\mathbf{r} = \mathbf{r}_{\rm ext}[0:P'-1] - j \mathbf{r}_{\rm ext}[P':2P'-1],
\eeq
where the indices inside the squared brackets represent the first and the last considered entries of the $\mathbf{r}_{\rm ext}$ vector, starting from entry zero.

The \mbox{non-zero} entries in the solution $\mathbf{r}$ reveal the presence of a path $p'$ with corresponding total delay $\tau_{p', \rm tot}$ with $p' \in \{0, \dots, P'-1\}$. 
Note that it would not be possible to separate the contributions in $\mathbf{r}$ as encoded in \eq{eq:r_model}. However, from \eq{eq:r_model} we know that the phase offset is the same across all the paths and we will leverage this fact in the following to sanitize the CFR phase.

Using the decomposition in \eq{eq:Hsplit}, the vector $\mathbf{r}$ found solving problem \textbf{P1}, and \eq{eq:T_matrix}, we are able to separate the contributions of the different paths in the CFR for each \mbox{sub-channel} $k$ by applying the following Hadamard product
\beq
    \label{eq:Hadamard_decomp}
	 \mathbf{X}_k = \mathbf{T}_k^T \circ \mathbf{r}.
\eeq
$\mathbf{X}_k$ can be rewritten by replacing the terms in \eq{eq:Hadamard_decomp}, as
\beq
\begin{split}
	 \mathbf{X}_k = &~ e^{- j2\pi k(\tau_{\rm SFO} + \tau_{\rm PDD})/T} e^{j(\phi_{\rm CFO} +\phi_{{\rm PPO}} + \phi_{{\rm PA}})} \times \\
	& \times 
	\begin{bmatrix}
	A_0 e^{-j2\pi (f_c+ k/T)\tau_0} \\
	\vdots \\
	 A_{P'-1} e^{-j2\pi (f_c + k/T) \tau_{P'-1}} 
	 \end{bmatrix}.
\end{split}
\eeq
At this point, we define $p^* \in \{0,\dots,P'-1\}$ as the position where $\mathbf{r}$ has the highest amplitude, which, in turn, is associated with the strongest path. Let $X_{k, p^*}$ be the $p^*$-th entry of $\mathbf{X}_k$. By multiplying $\mathbf{X}_k$ by the complex conjugate of $X_{k, p^*}$, we remove the phase components that are constant across all the paths, including the offset, obtaining
\beq \label{eq:reconstruct}
\begin{split}
	 \bar{\mathbf{X}}_k &= A_{p^*}e^{j2\pi (f_c+k/T)\tau_{p^*}}
	 \begin{bmatrix}
	A_0 e^{-j2\pi (f_c+ k/T)\tau_0} \\
	\vdots \\
	 A_{P'-1} e^{-j2\pi (f_c + k/T) \tau_{P'-1}} 
	 \end{bmatrix}.
\end{split}
\eeq
Hence, summing up the elements of $\bar{\mathbf{X}}_k$ (associated with the paths from $p'=0$ to $p'=P'-1$), we attain a CFR estimate for each of the \mbox{sub-channels}, where the phase offset is mitigated, as follows,
\beq \label{eq:hest}
\begin{split}
	\hat{H}_{k} &= \hat{A}_k(n) e^{j\hat{\phi}_k(n)} \\
	&= A_{p^*}\sum_{p'=0}^{P'-1} A_{p'} e^{-j2\pi (f_c + k/T)\hat{\tau}_{p'}}\\
	&\simeq A_{p^*}e^{j2\pi (f_c + k/T)\tau_{p^*}} H_k,
\end{split}
\eeq
where $\hat{\tau}_{p'} = \tau_{p'}-\tau_{p^*}$. The last relation -- that links $\hat{H}_{k}$ with $H_k$ -- requires an approximate equality sign as, in general, $P' \neq P$ and the paths $p'\in \{0, \dots, P'-1\}$ obtained by solving the compressive sensing reconstruction problem in $\mathbf{P1}$ may deviate from the actual $p \in \{0, \dots, P-1\}$ ones.
Note that $\hat{H}_{k}$ of \eq{eq:hest} represents the CFR estimate for sub-channel $k$, where the phase offset has been removed and the contribution of each path is modulated according to the amplitude and phase of path $p^*$. \vspace{-0.3cm}

\subsection{Doppler trace computation}\label{subsec:doppler_computation}
\begin{figure}[t!]
	\begin{center}   
	\begin{minipage}[]{0.49\columnwidth}
		\includegraphics[width=1\columnwidth]{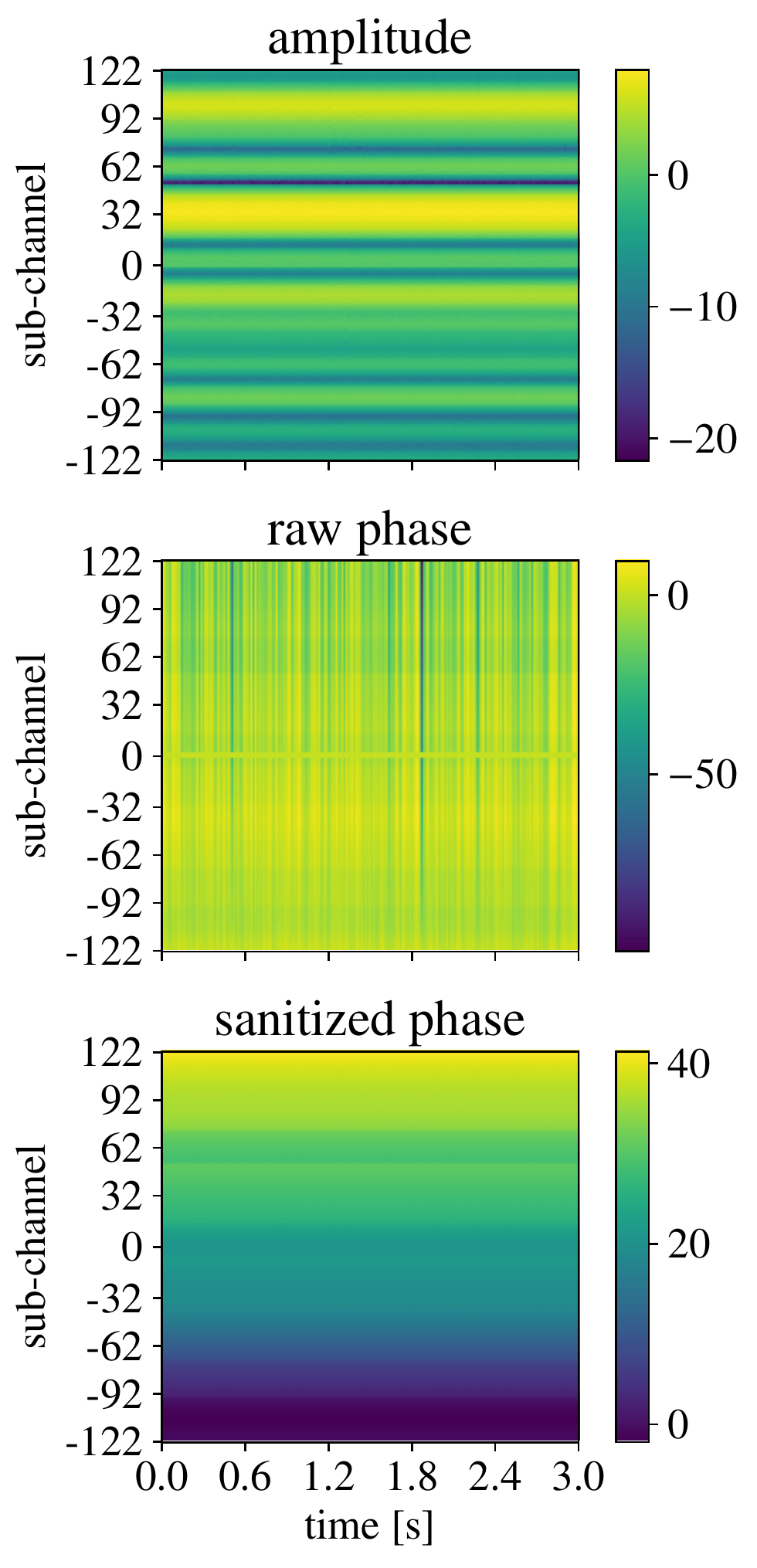}
	\end{minipage}
	\vspace{0.01\columnwidth}
	\begin{minipage}[]{0.49\columnwidth}
		\includegraphics[width=1.03\columnwidth]{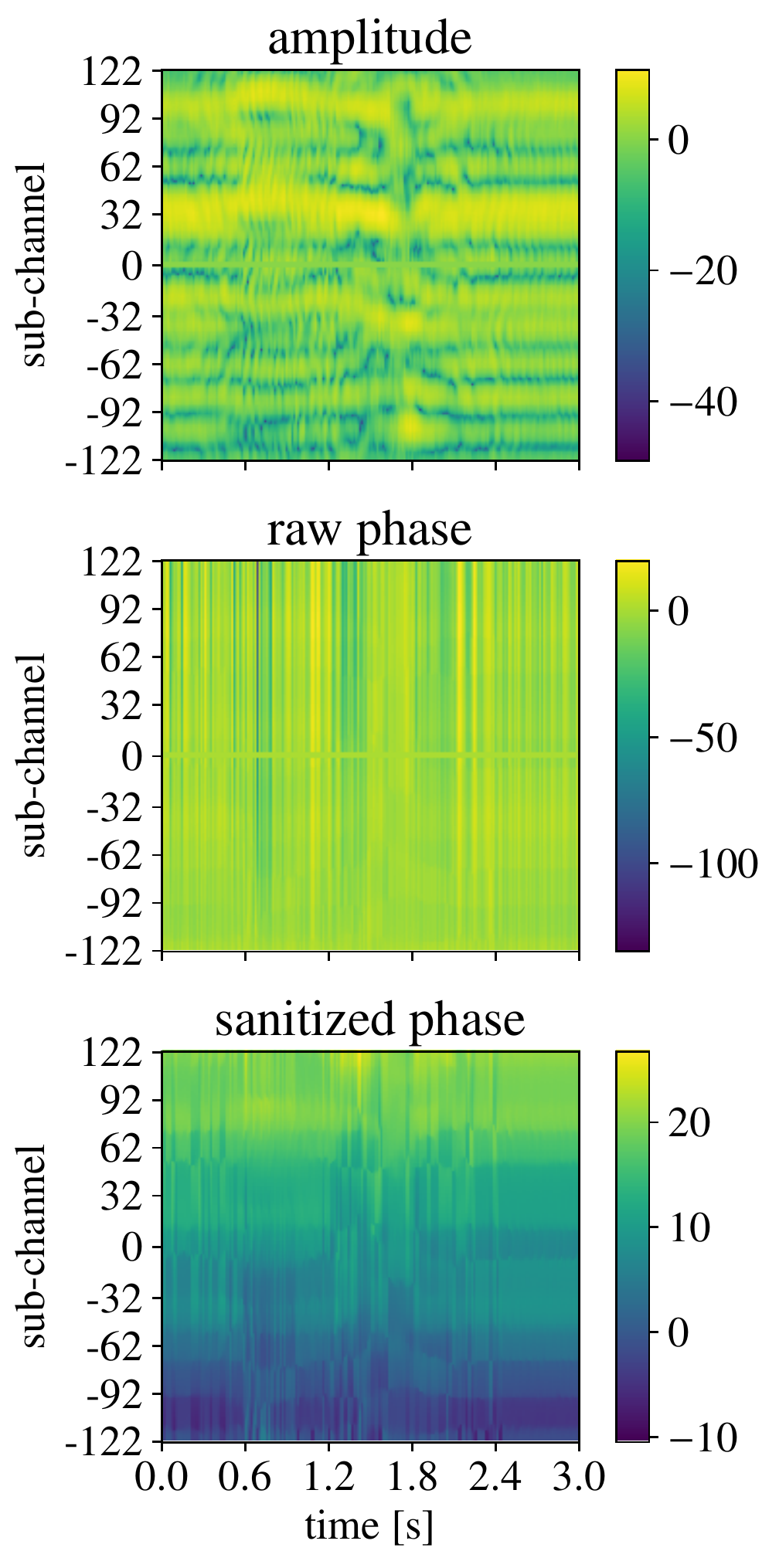}
	\end{minipage}
	\setlength\abovecaptionskip{-.1cm}
	\setlength\belowcaptionskip{-.5cm}
	\caption{Amplitude (in dB scale), raw and sanitized phases (unwrapped) of CFR data for an empty room (left plots) and with a person running (right plots). Each trace is three seconds long and shows the behavior on each of the monitored sub-channels ($y$-axis). Note that CSI is not available on the three central \mbox{sub-channels}, see Section~\ref{sec:setup} for details.} \label{fig:phase}
	\end{center}
\end{figure}

\begin{figure}[t!]
	\begin{center}   
	\begin{minipage}[]{0.49\columnwidth}
		\includegraphics[width=1.\columnwidth]{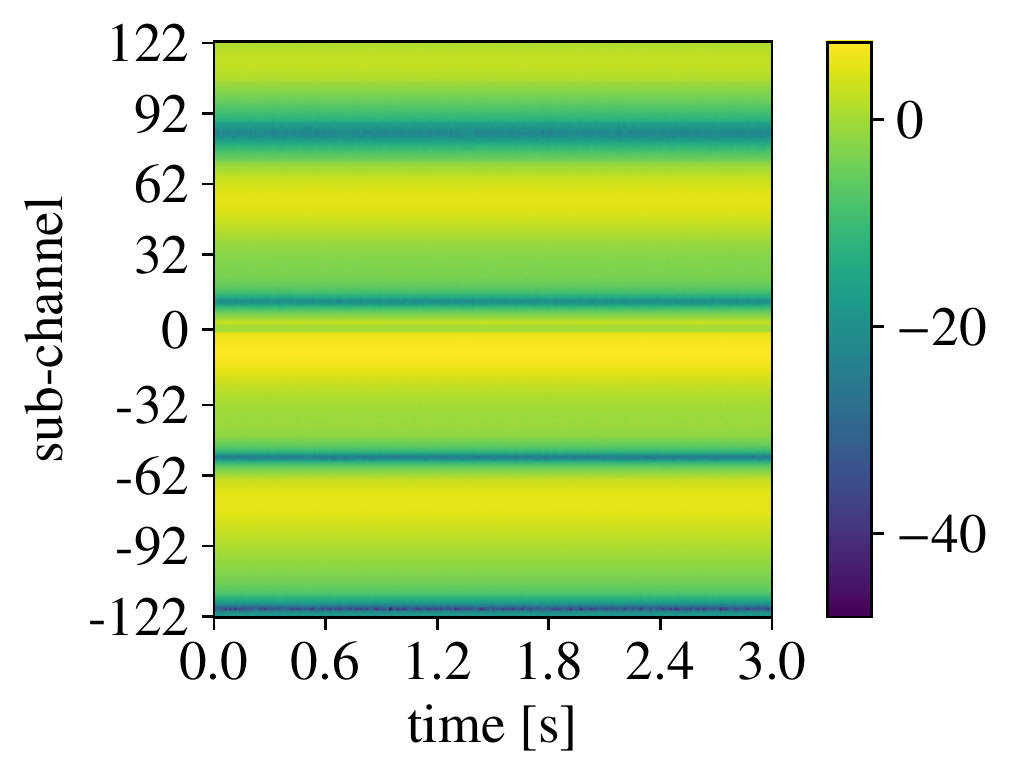}
	\end{minipage}
	\vspace{0.01\columnwidth}
	\begin{minipage}[]{0.49\columnwidth}
		\includegraphics[width=1.\columnwidth]{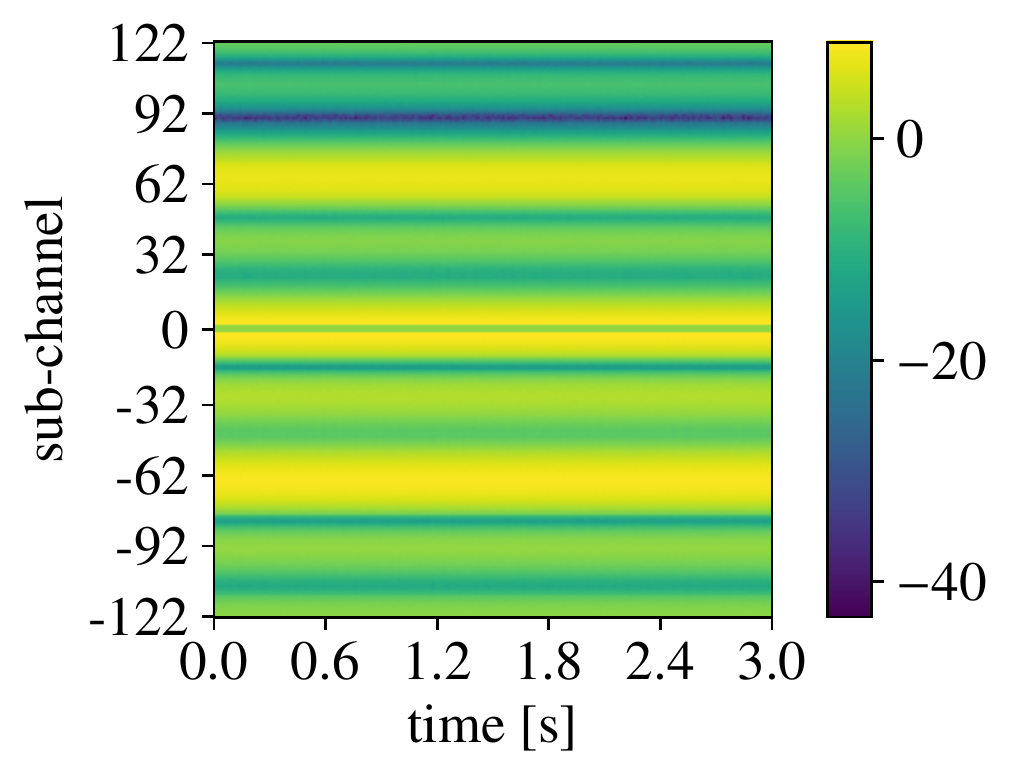}
	\end{minipage}
	\setlength\abovecaptionskip{-.1cm}
	\setlength\belowcaptionskip{-.6cm}
	\caption{CFR amplitude (dB scale) collected in the same empty room but in two different days. Environmental variations from one day to another change the multi-path CFR.} \label{fig:amplit}
	\end{center}
\end{figure}

\figurename~\ref{fig:phase} shows how the amplitude and the phase change during a time interval of $3$ seconds in two different scenarios, i.e., an empty room (on the left) and a room with a moving person (right plots). The presence of a human induces changes in the extracted channel parameters, which can be exploited by HAR algorithms. However, the CFR is affected by the whole indoor \mbox{multi-path} propagation environment and hence also accounts for the reflections from static objects. The reflected signals combine differently at each \mbox{sub-channel}, and the delays induce \mbox{sub-channel} specific phase shifts (see~\eq{eq:completeh_clean}). Such a behavior is {\it environment-specific} and is clearly identifiable in the amplitude plots of~\figurename\ref{fig:phase} (see the horizontal patterns in the figure). This fact does not allow developing robust algorithms for HAR that generalize across different environments, as the CFR is strongly affected by the room configuration itself (static objects, including walls, and the room shape). Also, even considering the same indoor space, slight changes in the position of the objects therein have a non-negligible effect on the measured CFR, thus making the HAR task more challenging. As an example, \figurename~\ref{fig:amplit} shows the CFR amplitude collected in two different days within the same empty room. Note that the presence of a person that remains still during the acquisition time modifies the CFR in a way that would be indistinguishable from changing the CFR of the same empty room by moving around static objects. We found that the micro-mobility of the chest and other body parts and, in turn, static poses, are hardly detectable using radio frequencies at sub-$6$~GHz and off-the-shelf IEEE 802.11ac technology, independently of the type of features extracted. Instead, static poses can be successfully identified and classified by leveraging the propagation behavior of radio waves at higher frequencies, e.g., in the millimeter-wave spectrum.

Due to these facts, in our framework we focus on the detection of dynamic activities and we exploit the Doppler effect to obtain effective features for environment-independent HAR. The Doppler effect corresponds to a shift in the signal phase measured at the receiver as the geometry of the multi-path propagation changes during a transmission event. The movements of the scatterers cause variations in the time needed by the signal to reach the receiver through each of the propagation paths. This reflects in a phase shift in the received OFDM signal and, in turn, in the CFR samples defined in \eq{eq:completeh_clean} (see Appendix~A for a description of the OFDM model). Specifically, considering path $p$ in $H_k(n)$, its associated delay, $\tau_{p}(n)$, can be expressed as the sum of two contributions. Let $\ell_{p}$ be the path length related to the initial position of the scattering point and $\Delta_p(n)$ be the delta caused by the movement of the point during the transmission period $nT_c$. We have
\beq \label{eq:tau_p}
	\tau_{p}(n) = \frac{\ell_{p}+\Delta_p(n)}{c},
\eeq
with
\beq
    \label{eq:integral_delta_p}
	\Delta_p(n)= -\int_{0}^{nT_c} v_p(x)\cos \alpha_p(x)dx,
\eeq
where $c$ is the speed of light, $v_p$ indicates the speed of scatterer $p$, and $\cos \alpha_p$ results from the combination of sinusoidal functions related to the angles of motion of the scatterer, and the angles of arrival/departure of the signal.
\begin{figure*}[t!]
	\begin{center}   
		\includegraphics[width=2.05\columnwidth]{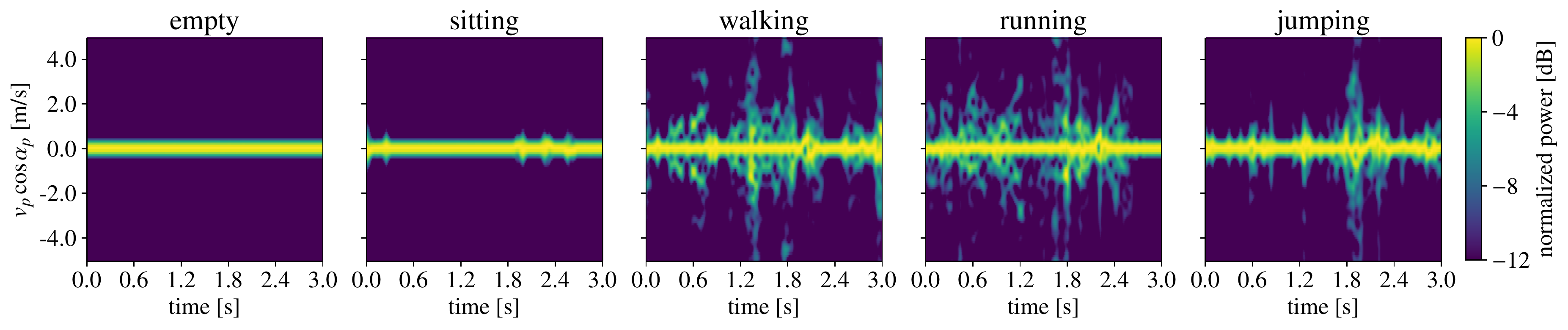}
		\setlength\abovecaptionskip{-.33cm}
	    \setlength\belowcaptionskip{-.62cm}
		\caption{Doppler spectrogram expressed as $v_p\cos\alpha_p$ for five different setups (from left to right: empty room, sitting, walking, running and jumping). The quantity can be positive or negative depending on the way the scattering points move, changing the geometry of the multi-path propagation channel.}
		\label{fig:doppler_traces}
	\end{center}
\end{figure*}

The \mbox{activity-related} movements of a human cause complex variations in the phase, as each body part acts as a scatterer moving at a specific velocity $v_p(x)$. This is revealed by the Doppler vector computed from the CFR samples through a \mbox{short-time} Fourier transform over $N$ subsequent channel estimates, i.e., during a {\it channel observation window}. These $N$ estimates are respectively acquired by extracting the CFR samples at the \mbox{Wi-Fi} monitor for $N$ subsequent packets, collected with sampling period $T_c$. The value of $N$ is selected so that the attenuation, the velocities, and the angles can be considered constant during the $i$-th observation window $[iN T_c, (i+1)N T_c]$, with $i \geq 0$. This allows us to rewrite $\Delta_{p}(n)$ (\eq{eq:integral_delta_p}) as
\beq\label{eq:deltA_p(t)}
	\Delta_{p}(n) = - v_{p}\cos \alpha_{p} nT_c.
\eeq
Our aim is to estimate the Doppler shift $v_{p}\cos \alpha_{p}$ from \eq{eq:hest} using \eq{eq:tau_p} and \eq{eq:deltA_p(t)}. However, note that, the sanitized phase expressed by~\eq{eq:hest} depends on $\hat{\tau}_p = \tau_p - \tau_{p^*}$, where $\tau_{p^*}$ is the propagation delay associated with the strongest path. Fortunately, we can reasonably assume that the strongest path, $p^*$, refers to a static component, thus  $\Delta_{p^*}(n) = 0$ and $\tau_{p^*} = \ell_{p^*}/c$, which is constant. It follows that the term depending on $p^*$ in \eq{eq:hest} does not alter the dynamic component of the signal, which is entirely contained in $H_k$ and, in turn, reveals the subject's movements. 
The reference static component is either a line-of-sight (LOS) or a non-line-of-sight (NLOS) path. Typically, in an indoor environment the signals propagate through several paths where at least walls contribute with static components even if the direct path between the transmitter and the receiver is blocked.
It descends that we are entitled to use \eq{eq:hest} to track the dynamic component in the Doppler domain. We do so by defining the $K\times N$ dimensional CFR matrix $\mathbf{H}_{i}$ associated with the $i$-th observation window, with entry $(k,n)$ referring to the CFR estimate from \eq{eq:hest} obtained for OFDM sub-channel $k$ and packet $n$ within the current window $i$, formally,
\beq \label{eq:H_matrix}
	 \mathbf{H}_{i} =
	 \begin{bmatrix}
	\hat{H}_{-K/2}(iN) & \dots & \hat{H}_{-K/2}((i+1)N -1) \\
	\vdots \\
	 \hat{H}_{K/2-1}(iN) & \dots & \hat{H}_{K/2-1}((i+1)N -1) 
	 \end{bmatrix}.
\eeq
A parallel can be built between matrix~\eq{eq:H_matrix} and the signal obtained by a \mbox{frequency-modulated} \mbox{continuous-wave} (FMCW) radar, where the {\it subcarrier} axis (row index) encodes the \textit{\mbox{fast-time}} component, while the {\it time} axis (column index, i.e., subsequent time estimates) serves as the \textit{\mbox{slow-time}} component~\cite{Pegoraro2020}. This analogy guides us on the transformation that is to be applied to the matrix to extract the desired Doppler information. Specifically, each element of the $N_D$-dimensional Doppler vector \mbox{$\mathbf{D}_i=[d_i(-N_D/2), \dots, d_i(N_D/2 - 1)]^T$} is obtained as
\beq \label{eq:doppler_trace}
d_i(u) = \sum_{k=-K/2}^{K/2-1} \left|\mathscr{F}\{\mathbf{H}_{i}\}(k, u) \right|^2,
\eeq
where $u \in \{-N_D/2, \dots, N_D/2 - 1\}$ is the Doppler index and $\mathscr{F}\{\cdot\}$ indicates the Fourier transform. 
\begin{proof}
The mathematical expression for $\mathscr{F}\{\mathbf{H}_{i}\}(k, u)$ is derived as follows.
According to the formulation in Section~\ref{sec:processing}, element $k,n$ of $\mathbf{H}_{i}$ is $\hat{H}_{k}(n)$. The models for the channel frequency response and the Doppler effect used in the computation are given in~\eq{eq:hest},~\eq{eq:tau_p} and~\eq{eq:deltA_p(t)}, while $W_{k}$ is the Hanning function, selected for the windowing operation. The term $e^{j2\pi (k\hat{v}_{p}\cos \hat{\alpha}_{p} nT_c/T)/c}$ is negligible and omitted in the final expression.
Note also that, to increase the velocity resolution, the signal can be \mbox{zero-padded} out to $N_D$ samples before applying the Fourier transform.
\beq \label{eq:fourier}
\begin{split}
\mathscr{F}\{\mathbf{H}_{i}\}(k, u) &= \sum_{n=0}^{N_D-1} \hat{H}_{k}(n)W_{k}(n) e^{-j2\pi n u/N_D} \\
&\simeq  A_{p^*} \sum_{p=0}^{P'-1} A_p e^{-j2\pi (f_c+k/T)\hat{\ell}_{p}/c} \times\\
&~~~~ \times \sum_{n=0}^{N_D-1}  W_{k}(n) e^{j2\pi n (f_c\hat{v}_{p}\cos \hat{\alpha}_{p} T_c/c - u/N_D)} .
\end{split}
\eeq
\end{proof}

By summing over the subcarrier axis $k$, in \eq{eq:doppler_trace}, we preclude the possibility of retrieving the length $\ell_{p}$ of path $p$ taking the maximum of the Fourier transform over $k$. However, by definition, $\ell_{p}$ is constant during a transmission slot and only depends on the position of the corresponding scattering point inside the room. In this work, we are interested in capturing the path variations ($\Delta_p(n)$) caused by a moving subject, thus the constant value $\ell_{p}$ is irrelevant to our HAR task.

The \mbox{non-zero} entries $u$ in the Doppler vector reveal the presence of a scatterer with associated velocity 
\beq \label{eq:vcos}
	v_{p}\cos \alpha_{p}  = \frac{uc}{f_cT_cN_D}.
\eeq
Note that, we do not need to estimate $v_p$ and $\cos\alpha_p$ separately. In fact, we are interested in finding a representative feature for the activity-related movements, i.e., we aim at revealing the dynamic component $\Delta_p(n)$ (\eq{eq:deltA_p(t)}) of the paths in the environment. It follows that the quantity in \eq{eq:vcos} is a good proxy to such dynamic components and is a good input feature for SHARP.
We recall that this feature is obtained considering time windows consisting of $N$ subsequent channel estimates each.
In the following, we refer to Doppler trace, or Doppler spectrogram, as the matrix obtained by stacking a number of subsequent Doppler vectors, computed for consecutive observation windows. 

As an example, in \figurename~\ref{fig:doppler_traces} we plot the Doppler spectrograms as a person performs each of the four considered activities inside a room, compared with an empty room case.
The spectrograms show the evolution of the quantity in \eq{eq:vcos} over \mbox{three-seconds} long observation windows and the colormap refers to the normalized amplitudes in dB. As expected, the Doppler trace related to the empty room only presents non-negligible power at the zero-velocity bin, revealing no movement in the environment. Instead, in the presence of a moving person the power is spread across different bins, reflecting the human-related multi-path changes.

We stress that the main idea behind SHARP is to perform environment and person independent activity recognition. To this end, we exploit the proprieties of the Doppler spectrogram discussed above combined with the learning architecture detailed in the next section. The framework is trained on a single scenario (i.e., same environment and person) and its robustness is assessed considering different test cases. Note that the framework is used at testing time without performing any retraining, to assess its performance when the day, the person and/or the environment change with respect to those used at training time.\vspace{-0.3cm}

\section{Learning architecture for HAR}\label{sec:learning}
SHARP is conceived to fully exploit the data gathered from a commercial \mbox{Wi-Fi} access point, i.e., using amplitude and phase information from the CFR, and combining this data for the available $N_{\rm ant}$ receiving antennas. Activity recognition is performed using $N_w$ subsequent channel estimates at a time, which amounts to monitoring the channel for $N_wT_c$~seconds. Note that this can be implemented in a sliding window fashion, thus updating the activity label at every new CSI sample.

The HAR algorithm consists of two steps. 
\begin{enumerate}[leftmargin=*]
    \item First, we compute the Doppler traces from the data collected at all the receiving antennas and we use them to obtain activity estimates though a neural-network based algorithm (Section~\ref{subsec:single}). Specifically, the machine learning based prediction chain that we build is {\it independently applied} to the data stream coming from each of the available antennas.
    \item As a result of the previous step, we obtain $N_{\rm ant}$ independent predictions, one per antenna, which are combined, in a second step, through a decision fusion method that leads to the final activity estimate (Section~\ref{subsec:fusion}).
\end{enumerate}
\begin{figure*}[t!]
	\begin{center}   
	\includegraphics[width=2.05\columnwidth]{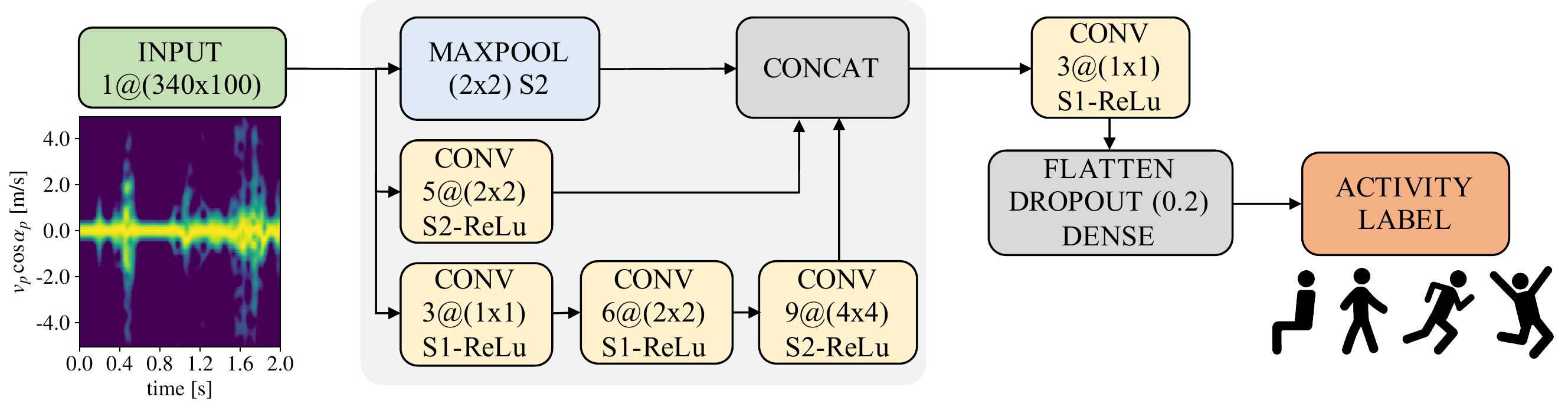}
	\setlength\abovecaptionskip{-.005cm}
	\setlength\belowcaptionskip{-.7cm}
	\caption{SHARP classification architecture for a single antenna system. The input of the network consists of $340$ (about two seconds) long Doppler traces with $100$ velocity bins. \mbox{[CONV maps@(kernel)]} blocks are convolutional filters with stride S$\#$, followed by ReLu activation functions.}\label{fig:pipeline}
	\end{center}
\end{figure*}
We remark that an alternative approach that combines the data from the antennas (at the input of the above decision chains), would also be possible. Such combined input data (the Doppler traces coming from the antennas) would then be fed to a neural network to classify the activity. We experimentally verified that this is not a viable method to obtain an environment-invariant classifier, as the classification result depends on the antenna ordering and, in turn, on the person's location inside the room. \vspace{-0.2cm}

\subsection{Activity classifier for the single antenna system}\label{subsec:single}
We framed the problem as a multi-class classification task with five classes (four user activities plus empty room), tackling it through a learning-based approach. The designed \mbox{neural-network} classifier takes as input the Doppler trace from a single antenna element, i.e., an $N_w\times N_D$-dimensional matrix obtained by stacking $N_w$ subsequent Doppler vectors, and returns the label of the activity being performed. The classification architecture for the single-antenna system is shown in \figurename~\ref{fig:pipeline}.

Firstly, the input is fed to a simplified Inception module with dimensional reduction. This block extracts significant features from the Doppler trace at several scales by using layers with different kernel sizes in a parallel fashion. The structure is inspired by the \textit{reduction block} proposed for the \mbox{Inception-v4} neural network in~\cite{Szegedy16}, and consists of three branches combining \mbox{max-pooling} (MAXPOOL) and convolutional (CONV) layers. Overall, the proposed neural network has $128,535$ parameters. In the interest of obtaining a lightweight model, we do not use the full 17 blocks \mbox{Inception-v4} network which consists of $43$ million parameters, but leads to negligible performance improvements.
The $N_w/2\times N_D/2$ dimensional feature maps obtained at the output of each branch of the Inception module are concatenated and passed to the following convolutional filter with $1 \times 1$ kernel, used to reduce the number of feature maps from $15$ to $3$. Hence, the output of such filter is flattened and passed to a fully connected (dense) layer with five output neurons, one for each activity class. The Dropout technique is used as a regularization strategy, randomly zeroing $20\%$ of the elements in the flatted vector preceding the dense layer. 
The classifier is trained in a supervised manner on data collected from a single indoor environment, using the \mbox{cross-entropy} loss function. The Doppler traces collected from the different antennas are used without any distinction among them, i.e., they are all added to a unique training set, without keeping track of the antenna that generated them. Using the trained architecture of \figurename~\ref{fig:pipeline}, each input trace is associated with the activity having the highest score in the output \mbox{five-dimensional} vector, referred to as {\it activity vector}. \vspace{-0.4cm}
\subsubsection{On the selection of the learning architecture}
The Inception module was selected among other candidate learning architectures as the subject's movements introduce both small and large scale variations in the Doppler spectrogram (see, e.g., \figurename~\ref{fig:doppler_traces}). As humans, we are able to detect and visually extract patterns at different scales, but this is not straightforward for a neural network, where each kernel captures features at a specific scale, depending on its size. The Inception module tries to mimic such a human-specific capability by simultaneously processing the input through different-sized kernels. As a consequence, in our case, the combination of the features extracted by an Inception module is expected to better represent the input than using a single kernel. 

An alternative approach to obtain such multi-scale features is to implement a deep neural network by stacking several convolutional layers. In this way, fine-grained features are extracted in the first steps and coarser characteristics are derived subsequently~\cite{goodfellow2016deep}. However, this approach leads to an architecture with a considerable number of trainable parameters which entails a longer training time and, more importantly, an increased memory to store the model. This limits the applicability of the method in the case of devices with scarce computing resources. As an alternative approach for resource constrained devices, the authors of~\cite{Huang2017Multi} propose the multi-scale dense-net (MSDNet). MSDNet consists of a convolutional architecture where layers of parallel sub-networks working at different scales are combined to reach scale invariance.
In this way, fine- and coarse-grained features are extracted at each layer of the combined network. In turn, each layer can output a reliable label for the classification problem and, at run-time, the processing can be stopped as soon as computing resources are exhausted. 
A similar approach is presented in~\cite{yang2020resolution} with the introduction of the resolution adaptive network (RANet). As for MSDNet, several parallel sub-networks operating at different scales are used. However, while in MSDNet the classification is always performed at the lowest scale, the RANet classifiers are placed both in low and high-resolution layers. The rationale behind this choice is that some ``easy'' input samples can be recognized directly at a higher scale without extracting coarse-grained features, thus reducing the computation burden.
Another scale-invariance approach is presented in~\cite{wang2019elastic}, where the authors introduce ELASTIC, a strategy that can be applied to any convolutional architecture. Specifically, down- and up-sampling functions are added as parallel branches at each layer of the network to change the spatial resolution of the extracted features. This allows obtaining a dynamic resolution structure where the network can activate branches with different resolutions at each layer, adapting the network to each input signal.

Overall, the approaches in~\cite{Huang2017Multi, yang2020resolution, wang2019elastic} are very useful in complex computer vision tasks. However, for our recognition problem, they would not lead to significant improvements at the cost of an increased model complexity. As our aim is to propose a lightweight approach that can be implemented in low-cost devices, we decided not to consider these more complex solutions.
\vspace{-0.2cm}

\subsection{Decision fusion for the multiple antenna system}\label{subsec:fusion}
At runtime, the trained classification engine from the single antenna system is independently applied to the Doppler stream gathered by  each antenna. This returns $N_{\rm ant}$ independent classification outcomes that are combined as we now explain. In detail, for each antenna we obtain a \mbox{five-dimensional} {\it activity vector} (the classifier output) and an {\it activity label}, corresponding to the largest element in the activity vector. When at least $N_{\rm ant}-1$ activity labels agree on a certain activity, there is a clear winner, and the Doppler trace is associated with that activity. Otherwise, an overall {\it decision vector} is computed by summing, element-wise, the $N_{\rm ant}$ activity vectors. The trace is then associated with the activity having the highest score in this decision vector.\vspace{-0.2cm}

\section{Experimental setup}\label{sec:setup}
In this section we present the experimental setup designed to train and validate SHARP. We first introduce the CSI extraction method (Section~\ref{subsec:nexmon}), and then present the measurement scenarios and campaigns selected for building the dataset (Section~\ref{subsec:dataset_1}). \vspace{-0.2cm}

\subsection{Nexmon extraction tool}\label{subsec:nexmon}
Although most commercial \mbox{Wi-Fi} chipsets can potentially generate CSI data, few manufacturers make this data available to developers and researchers, especially for modern chipsets. 
Hence, the majority of the \mbox{Wi-Fi}-based HAR works in the literature have used outdated chipsets, for which some CSI extraction tools have been developed over the years: widely used ones are~\cite{Halperin2011, Xie2015}, that target network interface cards implementing the IEEE 802.11n protocol. 
Recently, as part of the Nexmon project~\cite{Schulz2017, COOK, SHADOW}, a firmware patch allowing the extraction of CSI from specific Broadcom/Cypress \mbox{Wi-Fi} chipsets has been released. 

In the present work, we use the Nexmon CSI extraction tool presented in~\cite{Gringoli2019} to obtain CSI data from an Asus RT-AC86U IEEE 802.11ac \mbox{Wi-Fi} router. 
The extraction tool is compatible with the very-high-throughput mode, defined by IEEE 802.11ac, working with a total bandwidth of $80$~MHz. Each CSI sample results in complex-valued channel information from $242$ data sub-channels for each transmit-receive antennas pair. In our experiments, with one transmitter antenna and four at the monitoring device, each CSI sample corresponds to four vectors of $242$ complex values. Although the total number of sub-channels at $80$~MHz is $256$, each antenna vector has $242$ components as the CFR is only provided for data sub-channels, namely sub-channels whose indexes are $\{-122, \dots, -2\}$ and $\{2, \dots, 122\}$, i.e., no CFR value is provided for the control sub-channels. Moreover, the values returned by the tool on the sub-channels from $-63$ to $122$ need an inversion on the sign, probably due to hardware artifacts. 

In the next sections, the Asus router that estimates the CFR is referred to as \textit{monitor} device. The name reflects the status of the wireless interface of the router, that should be set in monitor mode to capture the packets sent over-the-air by other devices transmitting in its proximity.\vspace{-0.3cm}

\subsection{Dataset acquisition and organization}\label{subsec:dataset_1}
  
\begin{figure}[t!]
\centering
		\includegraphics[width=1\columnwidth]{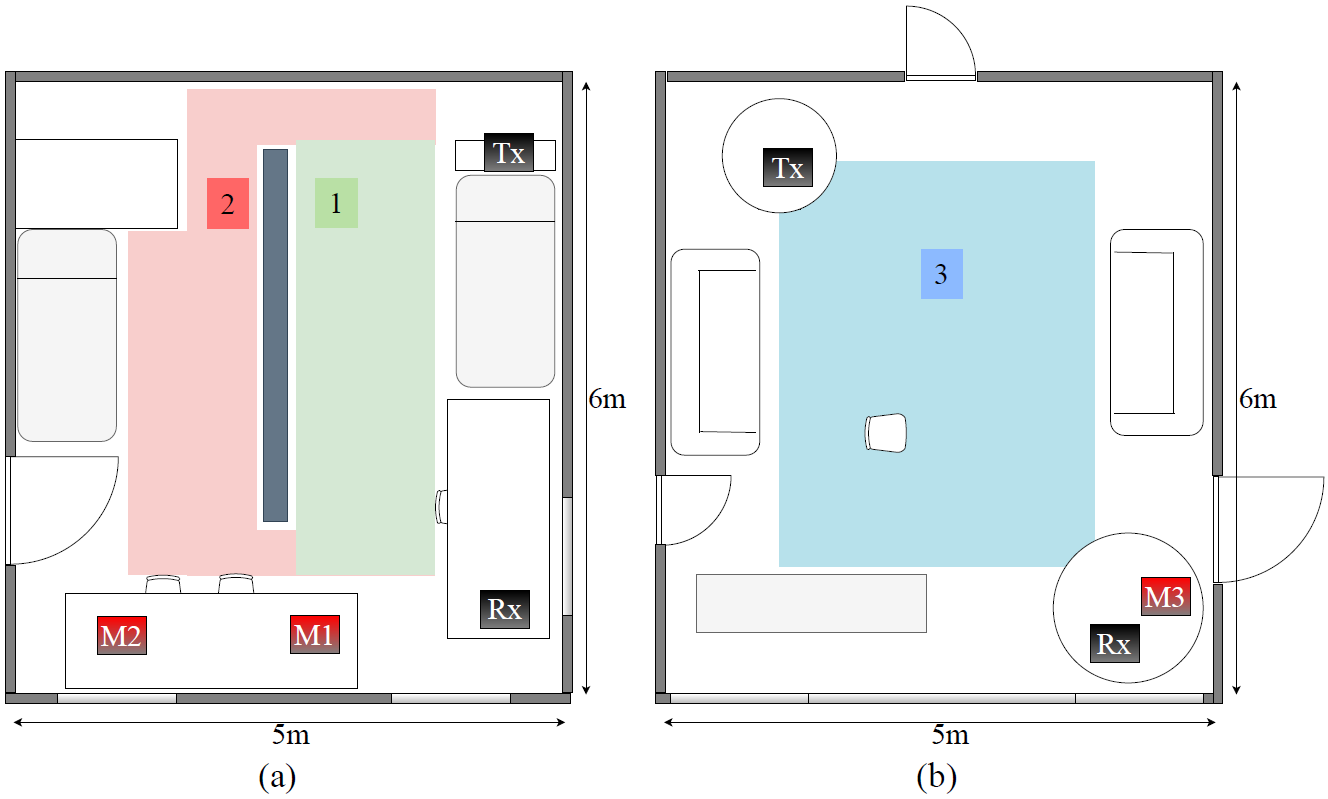}
		\setlength\abovecaptionskip{-.3cm}
		\setlength\belowcaptionskip{-.2cm}
		\caption{Monitored environment for (a) sets $S1$ - $S5$, bedroom with a bookcase on the middle; (b) set $S6$, living room. \texttt{Tx} and \texttt{Rx} denote the transmitter and the receiver, respectively. \texttt{Mj}, \texttt{j} $\in \{1, 2,3\}$, denotes the monitor station.}
		\label{fig:bedAndLive}
\end{figure}

\begin{figure}[t!]
\centering
		\includegraphics[width=0.9\columnwidth]{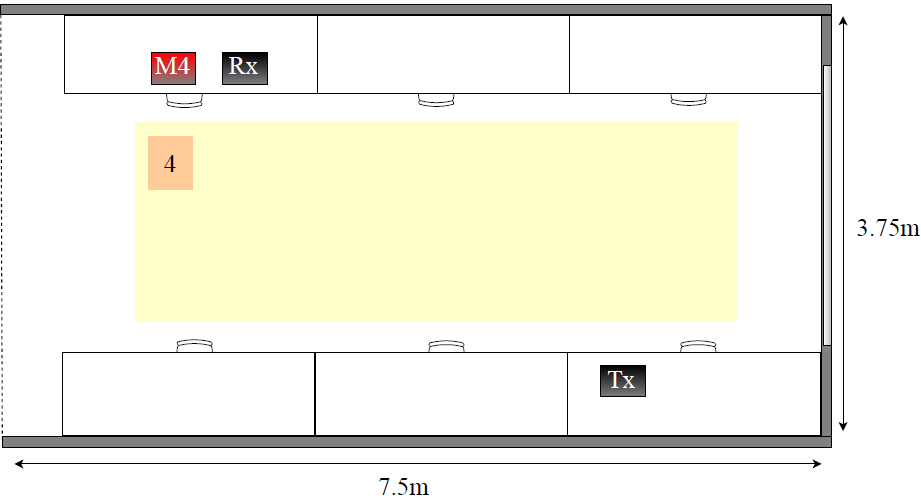}
		\setlength\abovecaptionskip{-.01cm}
		\setlength\belowcaptionskip{-.5cm}
		\caption{Monitored environment for set $S7$. The desks are fully of computers and monitors. \texttt{Tx} and \texttt{Rx} denote the transmitter and the receiver, respectively. \texttt{M4} denotes the monitor station.}
		\label{fig:lab}
\end{figure}

In order to  analyze the effects of different room geometries and static obstacles, we collected CSI samples in three different environments, i.e., a bedroom (\figurename~\ref{fig:bedAndLive}-a), a living room (\figurename~\ref{fig:bedAndLive}-b) and a University laboratory (\figurename~\ref{fig:lab}), where one person moves within the area.
Specifically, we obtained data from three volunteers (a male, \texttt{P1}, and two females, \texttt{P2}, \texttt{P3}) while they were walking or running around, jumping in place, or sitting somewhere in the room.
We recall that, during a \mbox{Wi-Fi} communication, the CSI samples are computed based on known packet preambles that allow estimating the channel conditions. CSI samples can be estimated by any device having access to the wireless channel where the communication is ongoing.
In our experimental setup, we considered a system made of two active terminals -- i.e., that exchange Wi-Fi traffic -- and one passive device configured in monitor mode. Such system reflects a real-world scenario where terminals (e.g., laptops and smartphones) transmit Wi-Fi packets that are leveraged by the monitor node as ``signals of opportunity'' to sense the surrounding environment through the computation of the CFR.
The monitor node is implemented on an Asus router equipped with $N_{\rm ant}=4$ antennas and running the Nexmon firmware (see Section~\ref{subsec:nexmon}).  
To generate the signals of opportunity, we set up a \mbox{Wi-Fi} transmission link using two Netgear X4S AC2600 \mbox{Wi-Fi} routers, equipped with Qualcomm Atheros chipsets and hereafter referred to as transmitter (\texttt{Tx}) and receiver (\texttt{Rx}) devices. Note that we used routers for experimental convenience: in a real setup Wi-Fi signals of opportunity can be generated by any other Wi-Fi-enabled device as it uses the spectrum for communications.
The packets are transmitted through a single antenna, by using a fixed modulation and coding scheme -- namely, MSC~$4$ -- by disabling frame aggregation, and by setting the source rate to $173$ packets per second. Since the Nexmon tool is configured for reading CSI samples on data packets only (i.e., by neglecting the acknowledgement frames sent by the receiver), a new channel estimate is generated every $T_c \simeq 6\times 10^{-3}$~s.

The proposed setup reflects a real-life situation where \mbox{Wi-Fi} traffic is already present in the environment, so we only need a monitor node to obtain the CSI traces from the signals of opportunity.
This is in line with the envisioned scenario for next-generation Wi-Fi routers implementing the upcoming IEEE 802.11bf standard~\cite{Restuccia2021}. The IEEE 802.11bf project started in September 2020 with the aim to empower \mbox{Wi-Fi} routers with sensing capabilities. The new version of the Wi-Fi standard will enable the joint provisioning of the communication and the sensing services to the users by introducing some modifications at the physical and medium access control layers of the protocol stack to fulfill the slightly different requirements of the two services. 
In this scenario, a Wi-Fi router will serve both as an access point for communication purposes and a monitor device to provide the sensing service by leveraging the ongoing Wi-Fi transmissions.
In our experimental setup, the monitor router provides the sensing functionality only as with the current physical and medium access control layer protocols the communication and the sensing services cannot be simultaneously provided to the users. However, our proposed sensing strategy will remain valid and effective for human activity recognition also once the two functionalities will be jointly supported by IEEE 802.11bf devices.

\figurename~\ref{fig:bedAndLive} and \figurename~\ref{fig:lab} display the positions of transmitter, receiver and monitor routers in the different environments.
The black boxes represent the transmitter (\texttt{Tx}) and the receiver (\texttt{Rx}), while the red boxes labeled with \texttt{Mj} (with \texttt{j}~$\in \{1, \dots, 4\}$) indicate the position of the monitor station in different measurement sets. 
The activities are performed in the areas identified by a color and a number.
Note that in the bedroom (\figurename~\ref{fig:bedAndLive}-a), the direct path between \texttt{Tx} and \texttt{M2} is occluded by the bookcase in the middle of the room (grey rectangle in the figure).

We performed several measurement campaigns and we grouped them into seven sets, $Sj$ with $j \in \{1,...,7\}$ each corresponding to a different triplet of environment-day-person.
For each set, Table~\ref{tab:configs} provides the position of the monitoring station (\texttt{Mj}), the area where the activities take place (identified by the color), and the person performing them ($\texttt{Pi}$).
We also include an indication of whether there is a direct path between the transmitter and the monitor stations.
The configuration for sets $S1$-$S2$ is the same apart from the day of measurement. Set $S1$ is used to train the model, while set $S2$ is used to test the generalization over different days.
In $S3$ we monitor the same environment as the previous sets, but in a different day and with a different person performing the activities.
For sets $S4$ and $S5$, the person is required to move in both area $1$ and $2$ of the bedroom depicted in~\figurename~\ref{fig:bedAndLive}-a. In these configurations, the direct path between the transmitter and the monitor is disturbed by the bookcase.
Set $S6$-$S7$ are collected in two different days and environments. $S7$ represents the more challenging situation, in which no element (room, day, person) is in common with the training set scenario.

Each measurement campaign involves $120$ seconds of data for each activity, plus an additional trace of $120$ seconds of data collected when the room is empty. Activities are repeated continuously by the volunteers during the trace acquisition time.  
The campaigns have been performed in different days through several months (April-December, $2020$ and January $2022$), resulting in a considerable time diversity. 
Overall, we collected nearly $120$ minutes of CSI data, consisting of the CFRs estimated at all the four antennas of the monitor station.\vspace{-0.3cm}
\begin{table}[t!]
\centering
\resizebox{\columnwidth}{!}{
\begin{tabular}{ccccc}
\toprule
\textbf{set} & \textbf{monitor position, \texttt{Mj}} & \textbf{person, \texttt{Pi}} & \textbf{direct path} & \textbf{type} \\ \midrule
$S1$ & \cellcolor{green!15}\texttt{M1} & \cellcolor{gray!10}\texttt{P1} &yes& train-test\\
$S2$ & \cellcolor{green!15}\texttt{M1} & \cellcolor{gray!10}\texttt{P1} &yes& test \\
$S3$ & \cellcolor{green!15}\texttt{M1} & \cellcolor{gray!40}\texttt{P2} &yes& test \\ 
$S4$ &  \diagfil{2.8cm}{red!20}{green!15}{\texttt{M2}} & \cellcolor{gray!10}\texttt{P1} &no& test\\
$S5$ & \diagfil{2.8cm}{red!20}{green!15}{\texttt{M2}} & \cellcolor{gray!40}\texttt{P2} &no& test \\
$S6$ & \cellcolor{cyan!25}\texttt{M3} & \cellcolor{gray!10}\texttt{P1} &yes& test \\
$S7$ & \cellcolor{yellow!35}\texttt{M4} & \cellcolor{gray!65}\texttt{P3} &yes&test\\
\bottomrule
\end{tabular}}
\caption{Measurement conditions. For each set $Sj$ we specify the position of the monitor station (\texttt{Mj}) (see \figurename~\ref{fig:bedAndLive} and \figurename~\ref{fig:lab}), the person (\texttt{Pi}) performing the activity, and the presence of a direct path between the transmitter and the monitor. The last column indicates whether the set is used for training SHARP or testing its performance.\label{tab:configs}}
\end{table}

\section{Experimental results}\label{sec:results}
SHARP has been tested in the scenarios detailed in Table~\ref{tab:configs}.
The adopted communication and processing parameters are summarized in Table~\ref{tab:parameters} and are presented in Section~\ref{subsec:parameters}. 
The performance is discussed in Section~\ref{subsec:accuracy}.
Our implementation of SHARP, along with the collected activity dataset, are available at~\cite{dataset-code}.\vspace{-0.3cm}
\begin{table}[t!]
\centering
\resizebox{\columnwidth}{!}{\begin{tabular}{lr}
\toprule
\multicolumn{2}{c}{\textbf{communication and processing parameters}} \\ \midrule
monitored channel & IEEE 802.11ac ch no. $42$ \\
OFDM sample duration, $T$ & $3.2 \times10^{-6}$~s \\
modulation and coding scheme & MCS 4 \\
channel estimates interval, $T_c$ & $\simeq 6\times10^{-3}$~s\\
no. OFDM sub-channels, $M$ & $256$ ($245$ used) \\
no. ch. estimates for Doppler computation, $N$ & $31$ \\
no. bins in a Doppler vector, $N_D$ & $100$ \\
no. Doppler vectors per window, $N_w$ & $340$ \\
no. monitoring antennas, $N_{\rm ant}$ & $4$\\
\bottomrule
\end{tabular}}
\setlength\belowcaptionskip{-.5cm}
\caption{Summary of parameters related to the communication setup, and the Doppler traces computation. \label{tab:parameters}}
\end{table}

\subsection{Pre-processing steps}\label{subsec:parameters}
We considered some pre-processing operations for extracting the features of interest from the dataset. 
For each CSI sample, we divided the CFR values by the mean amplitude over the $242$ monitored \mbox{sub-channels} to remove unwanted amplifications. 
Then, the phase sanitization algorithm presented in Section~\ref{subsec:sanitization} was applied, fixing $\lambda = 10^{-1}$. We reconstructed the CFR values on the three central \mbox{sub-channels} together with the other $242$ using \eq{eq:reconstruct}, thus obtaining a CFR complex-valued vector (amplitudes and phases) of $245$ components.
The Doppler vectors have been computed considering $N=31$ subsequent sanitized CFR samples. The velocity resolution was increased by zero-padding the signal out to $N_D = 100$ points before applying the Fourier transform. 
A threshold was used on the resulting Doppler vectors to remove noisy contributions with power smaller than $12$~dB.
Finally, the Doppler trace acting as input feature for the HAR system was built by stacking $N_w=340$ consecutive Doppler vectors. The Doppler vectors were generated for each CSI acquisition by using a sliding window mechanism. Therefore, the complete Doppler trace lasted roughly $340 \cdot T_c = 2$~s of measurements. \vspace{-0.2cm}

\subsection{SHARP training and performance assessment}\label{subsec:accuracy}
\begin{table}
\centering
\tabcolsep=0.11cm
\resizebox{\columnwidth}{!}{\begin{tabular}{c|c||c|c|c|c|c|c}
\multicolumn{2}{c}{}  & empty & sitting & walking & running & jumping & \textbf{mean} \\ \midrule
\multirow{2}{*}{$\bm S1$}&\textbf{accuracy ($\bm \%$)}& $100$ & $100$ & $100$ & $100$ & $97.93$ & $\mathbf{99.59}$ \\
&\textbf{F1-score}& $1$ & $1$ & $0.997$ & $0.996$ & $0.990$ & $\mathbf{0.996}$ \\ \midrule
\multirow{2}{*}{$\bm S2$}&\textbf{accuracy ($\bm \%$)}& $100$ & $100$ & $100$ & $99.28$ & $99.64$ & $\mathbf{99.78}$ \\
&\textbf{F1-score}& $1$ & $0.998$ & $0.999$ & $0.996$ & $0.998$ & $\mathbf{0.998}$ \\ \midrule
\multirow{2}{*}{$\bm S3$}&\textbf{accuracy ($\bm \%$)}& $100$ & $100$ & $96.64$ & $100$ & $100$ & $\mathbf{99.33}$  \\
&\textbf{F1-score}& $1$ & $1$ & $0.983$ & $0.984$ & $1$ & $\mathbf{0.993}$ \\ \midrule
\multirow{2}{*}{$\bm S4$}&\textbf{accuracy ($\bm \%$)}& $100$ & $100$ & $99.84$ & $93.27$ & $94.63$ & $\mathbf{97.55}$ \\
&\textbf{F1-score}& $1$ & $1$ & $0.965$ & $0.941$ & $0.972$ & $\mathbf{0.976}$ \\ \midrule
\multirow{2}{*}{$\bm S5$}&\textbf{accuracy ($\bm \%$)}& $100$ & $100$ & $98.58$ & $84.93$ & $99.45$ & $\mathbf{96.59}$ \\
&\textbf{F1-score}& $1$ & $1$ & $0.923$ & $0.911$ & $0.997$ & $\mathbf{0.966}$ \\ \midrule
\multirow{2}{*}{$\bm S6$}&\textbf{accuracy ($\bm \%$)}& $99.76$ & $100$ & $100$ & $100$ & $99.20$ & $\mathbf{99.79}$ \\
&\textbf{F1-score}& $0.999$ & $0.999$ & $0.999$ & $0.997$ & $0.996$ & $\mathbf{0.998}$ \\ \midrule
\multirow{2}{*}{$\bm S7$}&\textbf{accuracy ($\bm \%$)}& $100$ & $100$ & $81.24$ & $98.71$ & $100$ & $\mathbf{95.99}$ \\
&\textbf{F1-score}& $1$ & $1$ & $0.89$ & $0.909$ & $1$ & $\mathbf{0.960}$ \\
\bottomrule
\end{tabular}}
\setlength\belowcaptionskip{-0.6cm}
\caption{SHARP performance (decision fusion). The accuracy and the F1-score are reported for each of the five classes, along with their average value. \label{tab:results}}
\end{table}
The HAR learning algorithm presented in Section~\ref{sec:learning} has been trained by using the features extracted on set $S1$. Specifically, $60\%$ of the data in $S1$ makes up the training set, while the remaining $40\%$ is evenly split between the validation and the test sets. The other measurement sets, i.e., $S2$-$S7$, are only considered in the test phase.
The classification accuracy and the F1-score obtained on each of the test sets are reported in Table~\ref{tab:results}.
Note that, for $S1$, the evaluation is performed on the test set only, i.e., on new data never seen during training. 
Overall, the mean recognition accuracy is higher than $95\%$, reaching almost $100\%$ when the environment and location of the monitor node ($M1$) remain the same of the training data, regardless of day of measurement and the person performing the activity ($P1$ or $P2$). 
We still obtain a good accuracy, around $97\%$, when the direct path between the transmitter and the monitor stations is blocked (sets $S4$ and $S5$). In this situation, the running activity is sometimes confused with the walking one, as revealed by the F1-score associated with these classes. 
However, this is a reasonable limit of our approach, as the activities are recognized by considering the velocity of different parts of the human body (i.e., the scattering sources), which can be very similar when people walk or run in indoor spaces. 
We also believe that a small confusion between the running and walking activities is admissible in almost all the fields of application of a \mbox{Wi-Fi} sensing system. For example, in a residential scenario, it is more relevant to correctly distinguish between static and dynamic activities, rather than providing details about the type of movement.

\begin{figure}[t!]
\begin{center}
 \includegraphics[width=0.625\columnwidth]{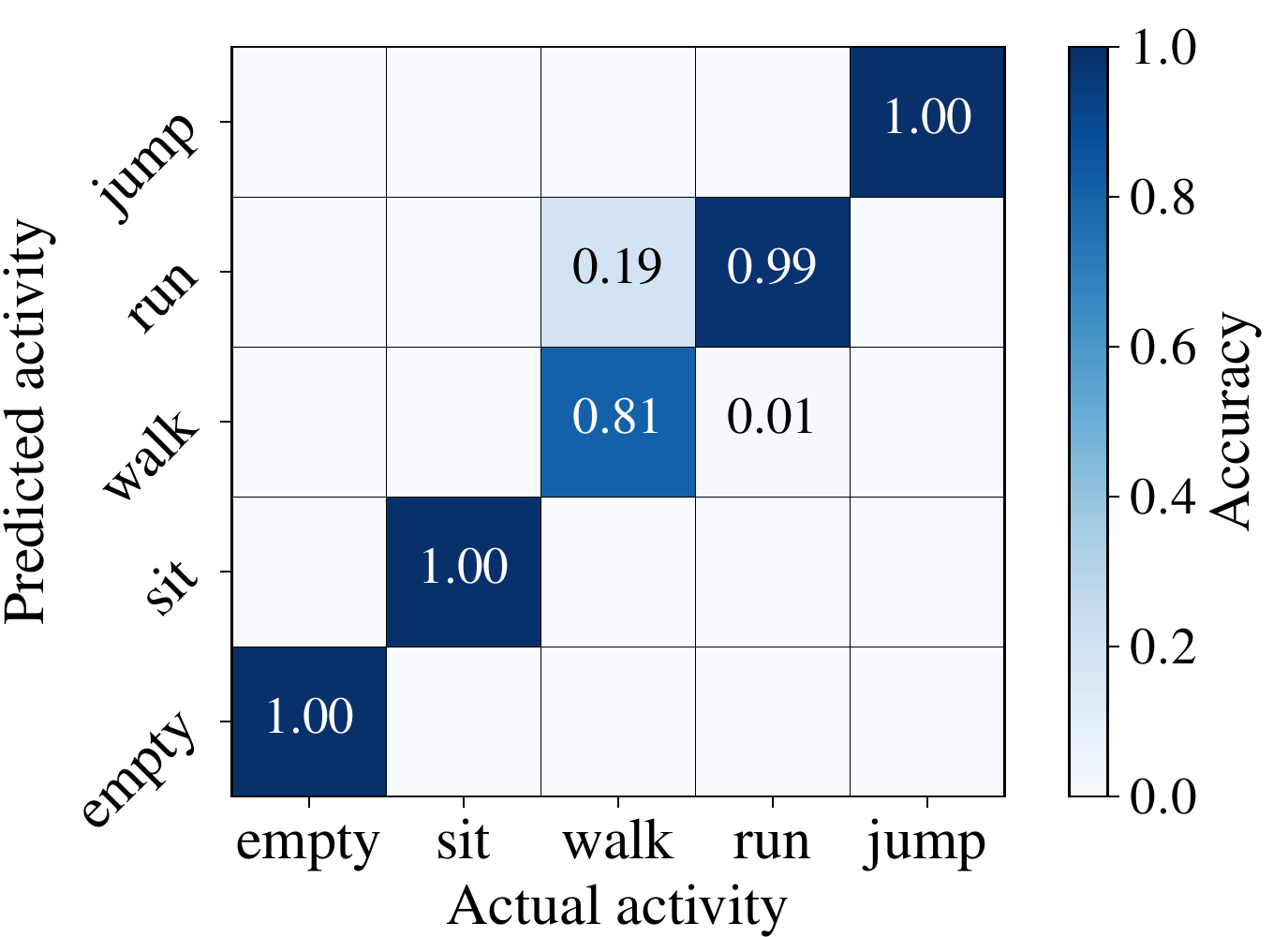} 
 \setlength\belowcaptionskip{-.7cm}
 \caption{SHARP normalized confusion matrix for test set $S7$ (worst case scenario). Environment, day and person change with respect to the training.}\label{fig:confusion_matrix}
\end{center}
\end{figure}

For testing the generality of our algorithm under different environments, we consider the measurement sets $S6$ and $S7$. When the person is the same of the training data, the average HAR accuracy approaches $100\%$ while it decreases to $95.99\%$ when both the person and the room change. Again, the lower accuracy is achieved for the walking activity, which is wrongly classified as the running one, as displayed in the normalized confusion matrix in \figurename~\ref{fig:confusion_matrix}.\vspace{-0.3cm}

\subsection{On the selection of the phase sanitization approach}\label{subsec:sanitization_comparison}
\label{sec:phase_sanitization_results}

\begin{figure*}[t!]
	\begin{center}   
	\includegraphics[width=2.05\columnwidth]{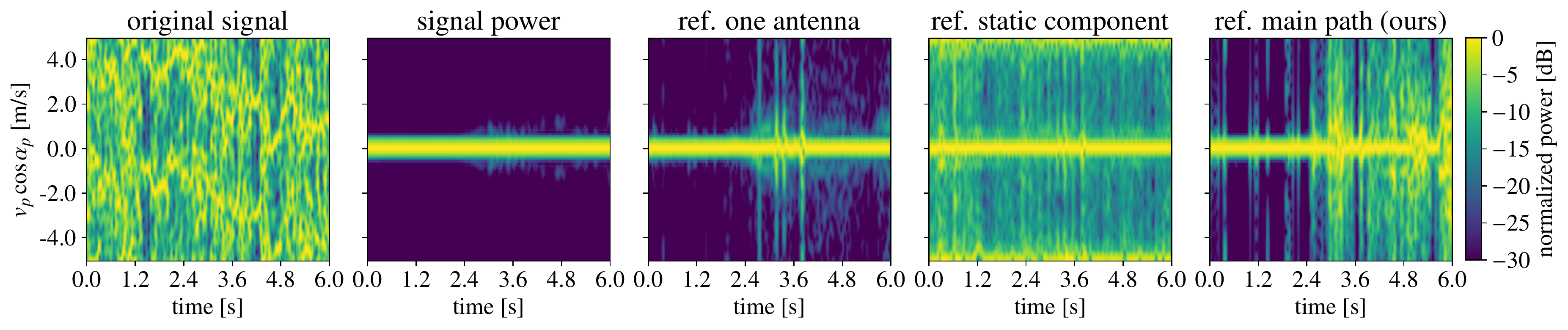}
	\setlength\abovecaptionskip{-.4cm}
	\setlength\belowcaptionskip{-.5cm}
	\caption{Doppler spectrograms computed from a six seconds long CFR signal, where the activity performed are sitting ($[0,3]$~s) and running ($[3,6]~s$). The first two subfigures show the Doppler extracted from the original signal (containing offsets) and the signal power. The following three plots represent the Doppler computed on the sanited signal -- i.e., obtained with three phase sanitization techniques -- labeled according to the reference (ref.) used.}\label{fig:sanitization_comparison}
	\end{center}
\end{figure*}
The removal of the phase offsets is necessary to obtain meaningful Doppler traces. The unwanted offsets, and in particular the CFO, disrupt the Doppler shift information contained in the CFR data, making its extraction impossible. Prior to developing our own phase sanitization method (see Section~\ref{subsec:sanitization}), we investigated existing phase sanitization techniques from the literature. The comparison between these strategies is not straightforward due to two main reasons. First, the target phase (i.e., the one without offsets) can be hardly obtained through a mathematical model, as the \mbox{multi-path} propagation is an \mbox{environment-specific} process that depends on many factors. One may think of evaluating the effectiveness of a phase sanitization algorithm by developing a simulator of the \mbox{Wi-Fi} channel, including the phase offsets introduced by the hardware imperfections. However, even with such simulator a fair comparison among the different strategies would be difficult, as they use different reference systems for the sanitized output signal. Specifically, the extensively adopted \mbox{conjugate-multiplication} between antennas (see, e.g.,~\cite{WangX2020, Zeng2018, Zhang2019, Zheng2019}), uses one of the antennas as the reference for the processed signal. The method in~\cite{Shi2020}, instead, refers the signal to its static component, while our proposed approach considers the strongest channel path as the reference.

As our main objective is to develop an environment-independent algorithm for HAR, we compare these phase sanitization approaches based on the quality of their features, i.e., by analyzing their respective Doppler spectrograms.
\figurename~\ref{fig:sanitization_comparison} shows the Doppler spectrogram computed from a $6$~second long CFR trace and its sanitized versions using the just mentioned techniques as a preprocessing step on the signal. In the considered time window, a person sits for the first $3$ seconds and then starts running inside the room. The ``original signal'' is clearly too noisy to be useful. A straightforward strategy, proposed in~\cite{Wang2015} to solve this issue, consists in computing the spectrogram from the power of the original signal that, naturally, is not affected by phase offsets. The result depicted in the second subfigure -- entitled ``signal power'' -- indicates that the sole amplitude information is not sufficient to obtain a clear spectrogram. The Doppler trace obtained with our sanitization algorithm (see Section~\ref{subsec:sanitization}) is referred to as ``ref. main path'', while the ``ref. one antenna'' and ``ref. static component'' respectively refer to the approaches in~\cite{Zheng2019} and~\cite{Shi2020}. 
The figure reveals that our proposed methodology allows obtaining a clearer Doppler spectrogram where the segments corresponding to the two different situations are more distinguishable.
The use of one antenna as the reference allows removing most of the phase offset contributions in the CFR, but it also removes some relevant information related to the environmental changes. This is due to the conjugate multiplication operation, which can lead to destructive interference between the signal to be cleaned and the reference. 
On the other hand, when the reference is the static component, the processed signal contains some noisy contributions which are probably associated with energy leakage between the static and the dynamic part. This prevents the correct estimation of the reference used for phase sanitization.
We experimentally verified that these previous algorithms do not lead to good HAR performance. These drawbacks led us to the development of our approach, detailed in the earlier Section~\ref{subsec:sanitization}.
A quantitative evaluation of the impact of the phase sanitization strategy is reported in Table~\ref{tab:sanitization_comparison}. Specifically, we assessed the accuracy and the F1-score of the classification algorithm (see Section~\ref{sec:learning}) using the Doppler spectrograms computed from the traces sanitized with the ``ref. one antenna'' strategy, and we compared the results with those obtained through our approach in Table~\ref{tab:results} (``ref. main path''). We selected the ``ref. one antenna'' as a benchmark for comparison as it returns the cleanest Doppler spectrograms among the other candidate approaches in the literature (see \figurename~\ref{fig:sanitization_comparison}). The results confirm the importance of the phase sanitization step for the extraction of informative Doppler traces that can serve as feature maps for HAR.
Note that the ``ref. one antenna'' is the most used approach in the literature. However, existing works on HAR only focus on the scenario here identified as $S1$. In $S1$, the performance obtained through the two sanitization methods are similar as the person and the environment remain the same at training and testing time. In this case, coarser Doppler spectrograms suffice to perform a correct classification of the activities. However, once the environment, the person or both change at testing time ($S2$-$S7$), the performance of the ``ref. one antenna'' sanitization method degrades, while the accuracy of our proposed strategy (``ref. main path'') remains above $95$\%.
\vspace{-0.3cm}
\begin{table}
\centering
\tabcolsep=0.11cm
\resizebox{1\columnwidth}{!}{\begin{tabular}{c|c||c|c|c|c|c|c|c}
\multicolumn{2}{c}{ } & {$\bm S1$} & {$\bm S2$} & {$\bm S3$} & {$\bm S4$} & {$\bm S5$} & {$\bm S6$} & {$\bm S7$} \\ \midrule
\multirow{2}{*}{\textbf{ref. one ant.}} & \textbf{accuracy ($\bm \%$)} & $99.23$ & $59.77$ & $75.99$ & $43.45$ & $39.20$ & $59.44$ & $60.31$ \\
& \textbf{F1-score}& $0.992$ & $0.563$ & $0.698$ & $0.323$ & $0.273$ & $0.556$ & $0.559$ \\
\midrule
\multirow{2}{*}{\textbf{ref. main path}} & \textbf{accuracy ($\bm \%$)} & $99.59$ & $99.78$ & $99.33$ & $97.55$ & $96.59$ & $99.79$ & $95.99$ \\
& \textbf{F1-score}& $0.996$ & $0.998$ & $0.993$ & $0.976$ & $0.966$ & $0.998$ & $0.960$ \\
\bottomrule
\end{tabular}}
\setlength\belowcaptionskip{-0.5cm}
\caption{Impact of using different phase sanitization methods for the estimation of the Doppler traces. Our approach (``ref. main path'') is here compared with the most adopted approach in the literature (``ref. one antenna'').}\label{tab:sanitization_comparison}
\end{table}

\subsection{On the selection of the fusion approach}\label{subsec:accuracy_data_fusion}
SHARP relies on the decision fusion strategy detailed in Section~\ref{subsec:fusion}. As an alternative, one may implement a data fusion approach. This consists in merging the data from the four antennas at the input of the neural network instead of combining the outputs of the classification performed on the single antennas. Hence, the $N_w\times N_D\times N_{\rm ant}$-dimensional input is forwarded through the neural network that returns the most probable human activity class. We implemented this approach by applying some slight modifications to the SHARP classification architecture of Section~\ref{subsec:single}. Specifically, we doubled the feature maps of the convolutional layer in the second branch of the simplified Inception block, while we used respectively $6$, $9$ and $12$ maps for the convolutional layers in the third branch. These adaptations are needed because of the higher dimensionality of the input matrix and lead to a total of $129,751$ trainable parameters. The network is trained by randomly swapping the order of the antennas in the $N_w\times N_D\times N_{\rm ant}$-dimensional input matrices to make the model independent of the antennas' positions along the channels dimension. Without the use of this strategy, the network would learn which antenna is the most important in the training scenario and will mostly base its classification on that one. However, the most significant antenna is likely to change when changing the environmental conditions thus leading to poorer classification performance.
The accuracy and F1-score of the data-fusion strategy averaged over the five considered activities are reported in Table~\ref{tab:results_datafusion}. The results show that the data fusion approach is less effective than the decision fusion one in recognizing the activity being performed by the subject, especially when the LOS is blocked -- $S4$ and $S5$ -- or the environment and subject both differ than in the training phase -- $S7$. This because by merging the four monitoring antennas’ data at the input of the neural network architecture, we force the algorithm to learn useful information from all the monitoring antennas. 
The shuffling of the antennas’ position at the input is required to grant independency from the antennas’ physical positioning. However, as a result, the signal captured by each monitor antenna has almost the same weight (importance) in the online classification outcome. This entails the fact that antennas that lead to a misclassifying of the activity have a considerable impact on the final activity classification as well. This drawback is bypassed by our proposed decision fusion approach of Section~\ref{subsec:fusion}: with it, we still take into account all of the antennas’ data when forwarding them through the neural network, one at a time. However, at the output, the decisions based on the single antennas are combined in a way that an antenna likely to lead to a misclassification -- as it is in discordance with the others -- is not considered for the final result.\vspace{-0.3cm}
\begin{table}
\centering
\tabcolsep=0.11cm
\resizebox{1\columnwidth}{!}{\begin{tabular}{c|c||c|c|c|c|c|c|c}
\multicolumn{2}{c}{ } & {$\bm S1$} & {$\bm S2$} & {$\bm S3$} & {$\bm S4$} & {$\bm S5$} & {$\bm S6$} & {$\bm S7$} \\ \midrule
\multirow{2}{*}{\textbf{data fus.}} & \textbf{accuracy ($\bm \%$)} & $99.80$ & $97.45$ & $89.43$ & $88.51$ & $81.68$ & $98.18$ & $88.02$ \\
& \textbf{F1-score}& $0.998$ & $0.966$ & $0.890$ & $0.884$ & $0.796$ & $0.982$ & $0.881$ \\
\midrule
\multirow{2}{*}{\textbf{decision fus.}} & \textbf{accuracy ($\bm \%$)} & $99.59$ & $99.78$ & $99.33$ & $97.55$ & $96.59$ & $99.79$ & $95.99$ \\
& \textbf{F1-score}& $0.996$ & $0.998$ & $0.993$ & $0.976$ & $0.966$ & $0.998$ & $0.960$ \\
\bottomrule
\end{tabular}}
\setlength\belowcaptionskip{-0.5cm}
\caption{Performance of the data fusion (fus.) approach compared to the decision fusion strategy in Section~\ref{subsec:fusion}. \label{tab:results_datafusion}}
\end{table}

\subsection{On the impact of the number of monitor antennas}\label{subsec:number_antennas}
\begin{figure*}[t!]
\begin{center}
    \subfigure[Impact of the number of monitor antennas combined by the decision fusion algorithm on the average recognition accuracy.]{\includegraphics[width=0.625\columnwidth]{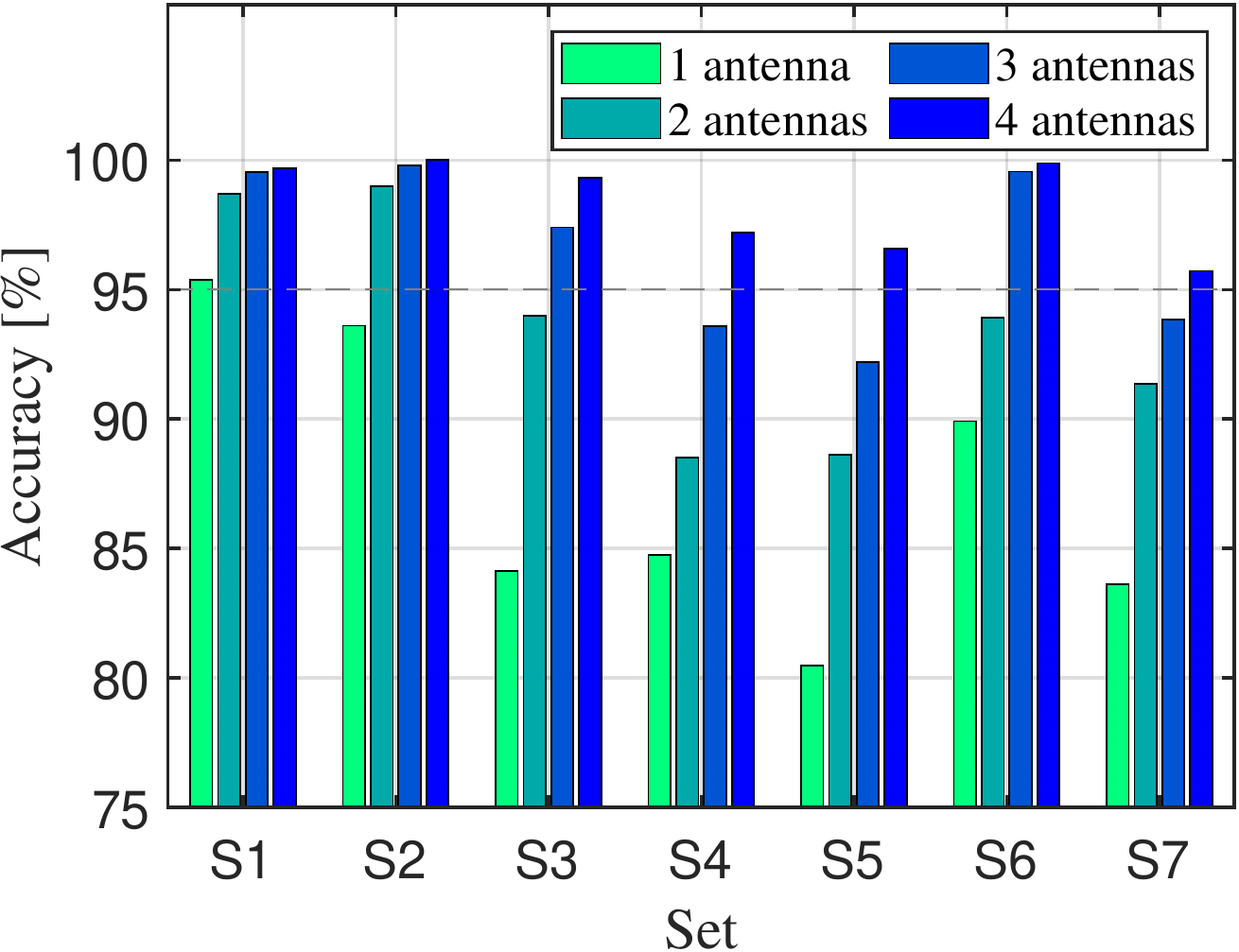}\label{fig:increase_number_antennas}}
	\hspace{0.2cm}
	\subfigure[Average accuracy of SHARP compared against the three considered approaches from the literature.]{\includegraphics[width=0.625\columnwidth]{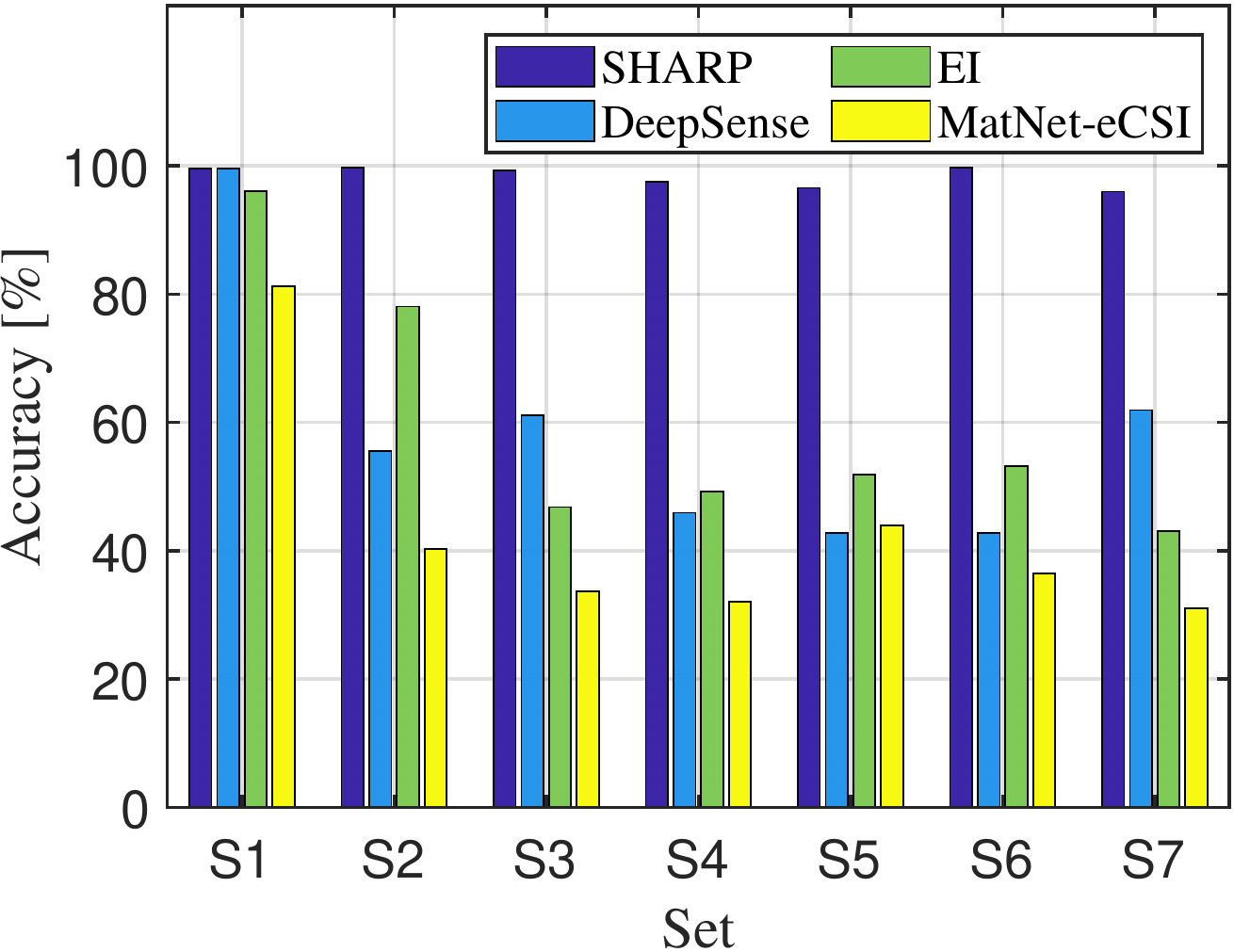}\label{fig:HAR_comparison_acc}}
	\hspace{0.2cm}
    \subfigure[Average F1-score of SHARP compared against the three considered approaches from the literature.]{\includegraphics[width=0.625\columnwidth]{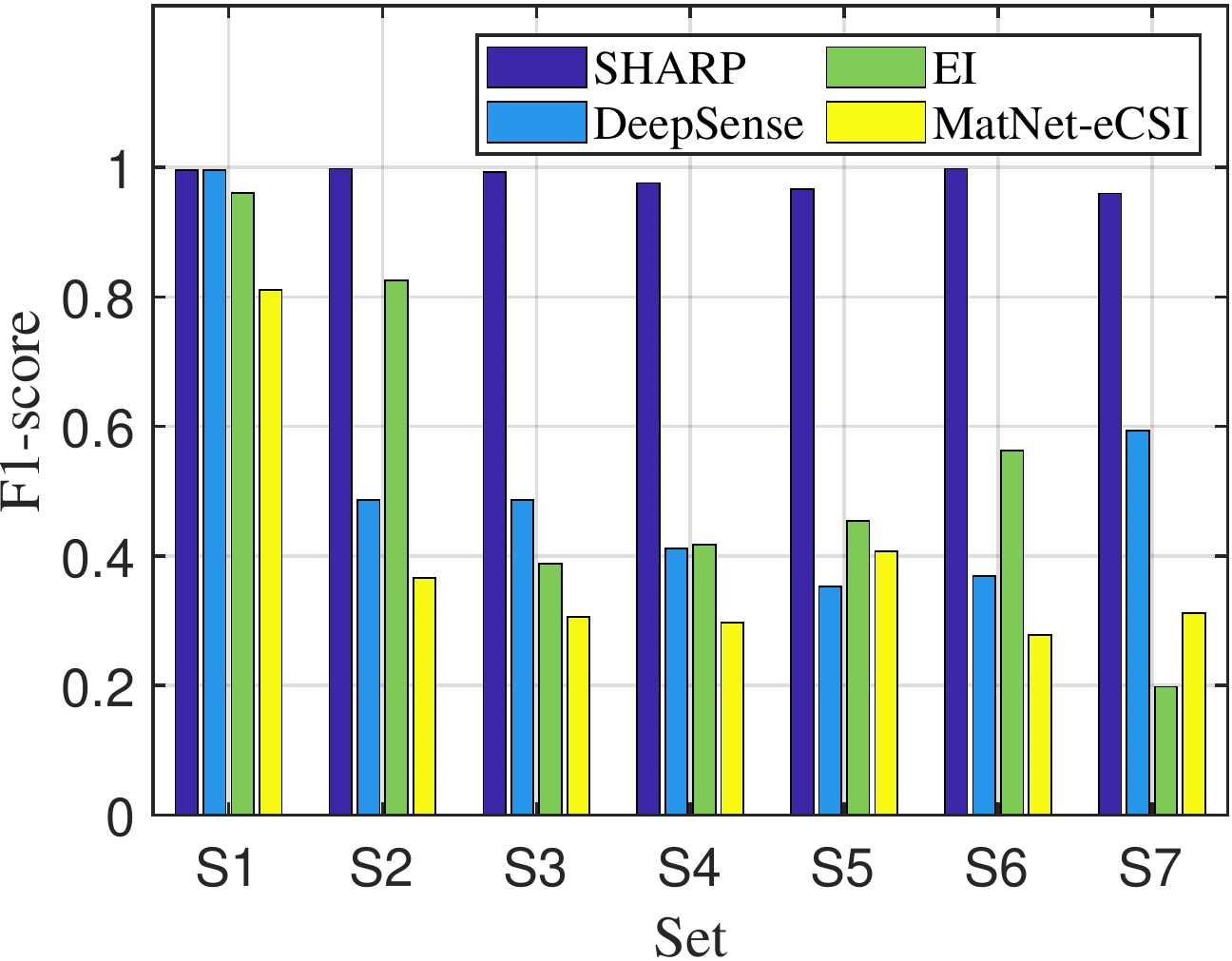}\label{fig:HAR_comparison_f1}}
    \setlength\abovecaptionskip{-.005cm}
	\setlength\belowcaptionskip{-.65cm}
    \caption{SHARP performance when changing the number of monitor antennas and comparison against three HAR algorithms from the literature -- namely DeepSense~\cite{Zou2018}, EI~\cite{TOWARDS}, and MatNet-eCSI~\cite{Shi2020} -- for all the configurations in Table~\ref{tab:configs}.}
		\label{fig:HAR_cm_and_comparison}
\end{center}
\end{figure*}

Here we evaluate the performance of SHARP when increasing the number of monitor antennas from $1$ to $4$. The neural network is trained as detailed in Section~\ref{subsec:single}. The classification output is obtained considering subsets of the four available antennas by applying the decision fusion algorithm of Section~\ref{subsec:fusion}. The results are reported in \figurename~\ref{fig:increase_number_antennas}. The activity recognition accuracy for $N_{\rm sub} \in \{1, \dots, 4\}$ antennas is obtained by averaging the metric over all activities and combinations of $N_{\rm sub}$ antennas. The results confirm that the higher the number of antennas, the higher the accuracy. This is associated with the amount of information that is collected for the classification. The accuracy is always above $80\%$ even with a single antenna. However, as confirmed by the results, the spatial diversity provided by the antenna array is beneficial for increasing the robustness of the system to changing environmental conditions. Having more antennas increases the possibility to sense environmental changes as the signals are collected at slightly different spatial positions.\vspace{-0.3cm}

\subsection{HAR algorithms performance comparison}\label{subsec:accuracy_comparison}
In Figs.~\ref{fig:HAR_comparison_acc}-\ref{fig:HAR_comparison_f1}, SHARP is compared against three HAR systems from the literature: DeepSense~\cite{Zou2018}, EI~\cite{TOWARDS} and \mbox{MatNet-eCSI}~\cite{Shi2020}. \textit{DeepSense} and \textit{EI} are used in place of the single antenna classifier described in Section~\ref{subsec:single} (i.e., before the decision fusion step) and rely on the sole CSI amplitude. \textit{MatNet-eCSI}, instead, considers phase and amplitude information, and takes as input their combination at all the antennas. The three approaches use different neural network architectures. 
In the interest of a fair comparison, we trained them with the same portion of the set $S1$ used to train SHARP. 
Note that \mbox{MatNet-eCSI} would require two training steps: the first performed on the training scenario, whereas the second on the test scenario, to fine tune the classification architecture to the new setup. For a fair comparison, in our evaluation we only considered the first step (using $S1$ data), i.e., without benefiting from additional data collected in the test scenarios. Specifically, we respectively used the training and the validation portions of $S1$ as the \textit{reference} and the \textit{target data} for the \mbox{MatNet-eCSI} neural network. For the EI algorithm, we considered the three different days containing the measurements for the training set $S1$ as the different domains for its adversarial learning framework. Then, for DeepSense and EI, the decision fusion approach of Section~\ref{subsec:fusion} is applied at runtime to combine the output of the classifiers operating on the single antenna data and obtain the final activity label. As \mbox{MatNet-eCSI} already takes the combination of features from all the antennas as input to its neural network classifier, the decision fusion is not performed for this scheme.

From Figs.~\ref{fig:HAR_comparison_acc}-\ref{fig:HAR_comparison_f1}, we observe that the performance of all the considered competing algorithms is in line with that of SHARP when tested on set $S1$, while it substantially degrades in all the other cases, even when data are collected in the same environment used for training, but on a different day (e.g., set $S2$).
These results demonstrate that the sole amplitude information (DeepSense and EI) or its combination with the phase (\mbox{MatNet-eCSI}) are insufficient for the classifiers to generalize well across different scenarios. 
Indeed, as shown and discussed in Section~\ref{sec:processing}, the CFR amplitude and its phase are \mbox{environment-dependent} and are affected by small scene variations over days, such as changes in the positioning of the monitor node, room obstacles, and objects. Conversely, SHARP generalizes across different days, environments and persons, by reliably distinguishing between static and dynamic situations. This is achieved by: i) exploiting the Doppler shift as the input feature for HAR, and ii) using the Inception module as the core of the learning-based classifier, to automatically extract coarse- and \mbox{fine-grained} details from the Doppler spectrograms.\vspace{-0.4cm}

\subsection{SHARP for smart-living applications} \label{subsec:more_activities}
In this section, we assess the performance of SHARP in recognizing a larger set of activities within a smart-living scenario. 
In addition to those already investigated, we added three additional human movements to the dataset, namely, standing still, sitting down/standing up and doing arm gymnastics, by retraining SHARP to distinguish them. The number of neurons in the last layer SHARP neural network classifier was changed to reflect the new number of classes. We experimentally verified that the standing-up and sitting-down movements produce an almost indistinguishable effect on the Doppler traces extracted from the Wi-Fi router. This is linked with the physical limitations given by the operational parameters (bandwidth and carrier frequency) of the used off-the-shelf Wi-Fi-based system. The obtained accuracy and F1-score are reported in Table~\ref{tab:results_smart} considering a single subject in LOS conditions. These results show that SHARP recognizes all the eight considered activities with high accuracy.
\begin{table}
\centering
\tabcolsep=0.11cm
\resizebox{\columnwidth}{!}{\begin{tabular}{c|c||c|c|c|c|c|c|c|c|c}
\multicolumn{2}{l}{}  & \multirow{2}{*}{empty}& \multirow{2}{*}{sitting} & \multirow{2}{*}{standing} & sit down & \multirow{2}{*}{walking} & \multirow{2}{*}{running} & \multirow{2}{*}{jumping} & \multirow{2}{*}{arm gym} & \multirow{2}{*}{\textbf{mean}} \\ 
\multicolumn{2}{l}{}  &  & & & stand up &  & & & & \\ \midrule
\multirow{2}{*}{$\bm S1$}&\textbf{acc. ($\bm \%$)}& $100$ & $100$ & $100$ & $100$ & $100$ & $100$ & $100$ & $96.02$ & $\mathbf{99.50}$ \\
&\textbf{F1-score}& $1$ & $1$ & $1$ & $0.994$ & $1$ & $1$ & $0.978$ & $0.980$ & $\mathbf{0.994}$ \\ \midrule
\multirow{2}{*}{$\bm S2$}&\textbf{acc. ($\bm \%$)}& $100$ & $100$ & $100$ & $100$ & $100$ & $99.84$ & $100$ & $94.56$ & $\mathbf{99.30}$ \\
&\textbf{F1-score}& $1$ & $1$ & $1$ & $0.994$ & $1$ & $0.999$ & $0.976$ & $0.972$ & $\mathbf{0.993}$ \\ \midrule
\multirow{2}{*}{$\bm S6$}&\textbf{acc. ($\bm \%$)}& $100$ & $100$ & $100$ & $100$ & $100$ & $73.16$ & $99.82$ & $90.22$ & $\mathbf{95.40}$ \\
&\textbf{F1-score}& $1$ & $1$ & $0.997$ & $0.960$ & $0.880$ & $0.844$ & $0.995$ & $0.949$ & $\mathbf{0.953}$ \\ 
\bottomrule
\end{tabular}}
\setlength\belowcaptionskip{-0.6cm}
\caption{SHARP performance with eight activities for scenarios $S1$, $S2$ and $S6$. The accuracy (acc.) and the F1-score are shown for each, along with their average value. \label{tab:results_smart}}
\end{table}

A remark is in order: The five activities considered in Table~\ref{tab:results} are reliably distinguished by SHARP across subjects, days and environments. This is due to the fact that the corresponding Doppler traces contain sufficiently separable dynamics, which can be reliably isolated. Instead, the three new activities that are considered in this section generate Doppler traces that are similar to those from the previous five. As such, their reliable separation requires looking at finer details, i.e., at subject-specific traits. This means that SHARP can reliably distinguish them across different days and environments, via a subject-specific training.\vspace{-0.3cm}

\section{Concluding remarks}\label{sec:conclusions}
In this work, SHARP, a novel \mbox{low-cost} system for human activity recognition (HAR) in indoor spaces has been presented and experimentally validated. This system analyzes \mbox{Wi-Fi} signals scattered into the environment and runs on COTS \mbox{Wi-Fi} routers, which are usually available in indoor spaces for communication purposes, thus making the deployment and maintenance of an \mbox{ad-hoc} sensing infrastructure unnecessary. The use of \mbox{Wi-Fi} routers as environmental sensors is enabled by the possibility of reliably estimating the wireless channel frequency response while receiving background data traffic.

In contrast with previous solutions, we implemented a {\it robust} HAR system that does not require complex and periodic calibrations, i.e., once trained, it can be used at \mbox{run-time} for a completely unseen setup (new environment and individuals). SHARP revolves around the idea of processing the CSI data gathered by a monitor access point to quantify the Doppler effect due to the presence of moving objects over time. Indeed, the Doppler shift describes the velocity of the scattering points (i.e., the human body parts), whose temporal trace depends on the specific activity and is neither affected by the environment geometry, nor by static objects. A learning based algorithm, working on the Doppler trace, has been proposed to distinguish among different activities.

Four coarse-grained human activities, i.e., sitting, walking, running and jumping are considered, together with the empty space recognition.
The robustness of SHARP has been challenged considering seven combinations of environments, access point locations, individuals and days. In each scenario, several measurement campaigns have been conducted, using a four-antennas IEEE~$802.11$ac router by Asus, operating on a frequency band of $80$~MHz. Experimental results show that our system achieves a very high accuracy in all the considered scenarios. In the most challenging situation, i.e., when the person and the environment change with respect to those used at training time, the average accuracy is $96\%$. Moreover, the system outperforms \mbox{state-of-art} methods based on the analysis of the CSI amplitude and phase. Small errors are observed only when working with unknown persons and when distinguishing between walking and running. These errors somehow represent the fundamental limits of the approach, because they are due to the behavioral differences among different persons, or to the movement similarities in walking and running activities. 
We experimentally verified that SHARP is effective even when considering additional activities with similar dynamics, namely standing, sitting down / standing up and doing arm gymnastics. The accuracy remains above $95\%$ when the performance is assessed in an environment that is different from the training one.

Future research avenues include the investigation of algorithms that work across differing hardware, at both the transmitter and the monitor access point, as well as the design of a new learning based algorithm that identifies multiple persons moving concurrently in the environment. Both extensions have important practical implications for real deployments and applications.\vspace{-0.35cm}

\section*{Acknowledgment}
This work has been supported, in part, by the Italian Ministry of Education, University and Research (MIUR) through the initiative ``Departments of Excellence'' (Law 232/2016) and by the European Union's Horizon 2020 programme under Grants No. 871249, project LOCUS. The views and opinions expressed in this work are those of the authors and do not necessarily reflect those of the funding institutions.\vspace{-0.3cm}

\appendices

\section{OFDM model for the \mbox{Wi-Fi} channel}\label{App:ofdm_deriv}
Next, we detail the main blocks of a \mbox{Wi-Fi} transmission chain, deriving the expression for the CFR in Eq. (2).

\subsection{Transmitted signal model}\label{sec:transmitter}
In this subsection, we summarize how orthogonal frequency-division multiplexing (OFDM) is implemented by an IEEE 802.11 transmitter. OFDM uses a large number of closely spaced orthogonal \mbox{sub-channels}, transmitted in parallel. Each \mbox{sub-channel} is modulated with a conventional digital modulation scheme (such as QPSK, 16QAM, etc.). 

The input bits are grouped and mapped onto source data symbols, which are complex numbers representing the modulation constellation points. Such constellation points are further grouped into $K$ elements each, i.e., the OFDM symbols. OFDM symbols are then fed to an IFFT block that transforms the data prior to transmitting it over the wireless channel, in parallel, over $K$ \mbox{sub-channels} with carriers spaced apart by $\Delta f = 1/T$~Hz ($T$ is the OFDM symbol time).

The duration of an OFDM symbol is $\bar{T}= T+ T_{\rm CP}$, where $T_{\rm CP}$ is the duration of the cyclic prefix, added to mitigate \mbox{inter-symbol} interference. Specifically, for IEEE 802.11ac the transmission bandwidth is $80$~MHz, the samples are clocked out at $80$~Msps, the number of \mbox{sub-channels} is $K=256$, $T=1/\Delta f=3.2$~$\mu$s (i.e., $\Delta f=312.5$~kHz), $T_{\rm CP}= 0.8$~$\mu$s, and, in turn, $\bar{T}= 4$~$\mu$s.

Let $\mathbf{a}_m= [a_{m, -K/2}, \dots, a_{m, K/2-1}]$ be $m$-th OFDM symbol, where $a_{m, k}$ is the $k$-th OFDM sample. After digital to analog conversion, the baseband OFDM signal for the $m$-th symbol is
\beq
	x_m(t) = \sum_{k=-K/2}^{K/2-1} a_{m, k} e^{j2\pi kt/T},
\eeq
where $k  \in \{-K/2, \dots, K/2 - 1\}$ is the sub-channel index. Considering $M$ subsequent blocks, the baseband signal is
\beq \label{eq:baseband}
	x(t) = \sum_{m=0}^{M-1}x_m(t)\xi (t-m\bar{T}),
\eeq
with 
\beq 
	\xi (t) = 
	\begin{cases}
	1 ~{\rm if}~ t \in [-T_{\rm CP}- T K/2, T K/2]\\
	0 ~{\rm otherwise}
	\end{cases},
\eeq
and the signal transmitted over the \mbox{Wi-Fi} channel is obtained by upconverting $x(t)$ to the carrier frequency $f_c$,
\beq
	s_{\rm tx}(t) = e^{j2\pi f_c t} x(t).
\eeq

\subsection{Received signal model}\label{sec:receiver}
At each receiver antenna, $P$ signal copies are collected, due to the scatterers that the signal $s_{\rm tx}(t)$ encounters (multi-path propagation). Each path $p$ is characterized by an attenuation $A_{p}(t)$ and a delay $\tau_{p}(t)$. Neglecting the additive white Gaussian noise, the received signal $s_{\rm rx}(t)$ is written as
\beq
\begin{split}\label{eq:srx}
	s_{{\rm rx}}(t) &= \sum_{p=0}^{P-1}A_{p}(t) s_{\rm tx}(t - \tau_{p}(t)) \\
	&= e^{j2\pi f_c t}\sum_{p=0}^{P-1}A_{p}(t) e^{-j2\pi f_c \tau_{p}(t)}x(t - \tau_{p}(t)),
\end{split}
\eeq
and its baseband representation $y(t)$ is expressed as,
\beq
\begin{split}\label{eq:ys}
	y(t) &= s_{{\rm rx}}(t) e^{-j2\pi f_c t}.
\end{split}
\eeq

A rectangular window $[m\bar{T}, m\bar{T} + T]$ is used at the receiver to collect and decode the information carried by an OFDM symbol at a time. Without loss of generality, we assume $m=0$ and hence we omit such index from the following equations. The transmitted symbol $a_{k}$ is recovered by computing the Fourier transform of the signal in the received window:
\beq \label{eq:proof_ak}
\begin{split}
	\hat{a}_{k} & = \int_{\bar{T}}^{\bar{T}+T} y(t)e^{-j2\pi kt/T} dt\\
	&= \sum_{p=0}^{P-1} A_{p} e^{-j2\pi f_c \tau_{p}} \sum_{b=-K/2}^{K/2} a_{b} e^{-j2\pi b\tau_{p}/T} \times \\ 
	&~~~~ \times\int_{\bar{T}}^{\bar{T}+T} e^{j2\pi (b-k) t/T}dt\\
	&= a_{k}T \sum_{p=0}^{P-1} A_{p} e^{-j2\pi (f_c + k/T) \tau_{p}},
\end{split}
\eeq
where we consider $A_{p}$ and $\tau_{p}$ constant over an integration interval $T$. The sum in the last line of~\eq{eq:proof_ak} corresponds to the frequency response of the \mbox{Wi-Fi} channel,
\beq \label{eq:estimates}
	H_{k} = \sum_{p=0}^{P-1} A_{p} e^{-j2\pi (f_c + k/T)\tau_{p}},
\eeq
that is estimated based on the known preamble symbols.
In \eq{eq:proof_ak}, we consider that the path attenuation and delay remain constant over each window, i.e., $A_{p}(t) = A_{p}$ and $\tau_{p}(t) = \tau_{p}$. Also, exchanging the order of integration and summation is legitimate as we deal with finite quantities, and we used $\int_{\bar{T}}^{\bar{T}+T} e^{j2\pi (b-k)t/T} dt = 0$ if $k \neq b$. 

\subsection{Phase offsets in the \mbox{Wi-Fi} channel estimates}\label{sec:channel_est}
Hardware artifacts make the CFR gathered from \mbox{Wi-Fi} devices slightly deviate from the model in \eq{eq:estimates}. These artifacts introduce offsets (rotation errors) in the phase information, among which the most significant are~\cite{Zhu2018, Keerativoranan2018}:
\bit
	\item \textit{carrier frequency offset (CFO)}, due to the difference between the carrier frequency of the transmitted signal and the one measured at the receiver. The CFO is only partially compensated for at the receiver~\cite{Chen2019}.
	\item \textit{sampling frequency offset (SFO)}, due to the imperfect synchronization of the clocks between transmitter and receiver.
	\item \textit{packet detection delay (PDD)}, due to the time required to recover the transmitted modulated symbols from the received signal~\cite{Chen2019}.
	\item \textit{phase-locked loop phase offset (PPO)}, due to the \mbox{phase-locked} loop (PLL), the entity responsible for randomly generating the initial phase at the transmitter. 
	\item \textit{phase ambiguity (PA)}, due to the phase difference (multiples of $\pi$) between the antennas, that in static conditions should remain constant.
\eit
Considering these contributions, the complete expression for the phase of the $p$-th path in the received signal is
\beq \label{eq:completephase}
\begin{split}
	\bar{\phi}_{p, k} =&~ -2\pi (f_c + k/T)\tau_{p} + \phi_{\rm CFO} \\
		&~ - 2\pi k(\tau_{{\rm SFO}} + \tau_{{\rm PDD}})/T \\
		&~ + \phi_{{\rm PPO}} + \phi_{{\rm PA}}.
\end{split}
\eeq

Note that, while the CFO, SFO and PDD contributions take the same value across different antennas, the initial PLL phase (PPO) and PA are antenna specific~\cite{Zubow2020, Zhang2020}. 

\bibliographystyle{IEEEtran}
\bibliography{IEEEabrv,biblio}
\end{document}